\documentclass[a4paper,11pt]{article}

\usepackage{jcappub} 

\usepackage[T1]{fontenc} 

\usepackage{hyperref}
\usepackage{amssymb}
\usepackage{bm}
\usepackage{dcolumn}
\usepackage{amssymb}
\usepackage{amsmath}
\usepackage{graphicx}
\usepackage{booktabs}
\usepackage{multirow}
\usepackage{setspace}
\usepackage{array}
\usepackage{threeparttable}
\usepackage{txfonts}
\usepackage{subfig}

\title{\boldmath Regularity of high energy photon events from gamma ray bursts\footnote{This paper has been published as H.Xu and B.-Q.Ma, JCAP 1801 (2018) 050.}}

\author[a]{Haowei Xu,}
\author[a,b,c,d]{Bo-Qiang Ma\footnote{Corresponding author at: School of Physics, Peking University, Beijing 100871, China. Email address: mabq@pku.edu.cn}}

\affiliation[a]{School of Physics and State Key Laboratory of Nuclear Physics and
Technology, Peking University, Beijing 100871, China}
\affiliation[b]{Collaborative Innovation Center of Quantum Matter, Beijing, China}
\affiliation[c]{Center for High Energy Physics, Peking University, Beijing 100871, China}
\affiliation[d]{Center for History and Philosophy of Science, Peking University, Beijing 100871, China}

\abstract{The effect of Quantum Gravity (QG) may bring a tiny light speed variation as $v(E)=c(1-E/E_{\rm LV})$, where $E$ is the photon energy and $E_{\rm LV}$ is a Lorentz violation scale. A remarkable regularity was suggested in previous studies to look for the light speed variation from high energy photon events of Gamma Ray Bursts (GRBs). We provide a general analysis on the data of 25 bright GRBs observed by the Fermi Gamma-ray Space Telescope (FGST). Such method allows a completed scan over all possibilities in a more clean and impartial way without any bias compared to previous intuitive analysis. The results show that with the increase in the intrinsic energies of photons, such regularity truly emerges and gradually becomes significant. For photons with intrinsic energies higher than 40~GeV, the regularity exists at a significance of 3-5~$\sigma$ with $E_{\rm LV}=3.6\times 10^{17}~\rm GeV$ determined by the GRB data.}

\begin{document}
\maketitle
\flushbottom

\section{Main Content}

The speed of light is assumed to be a constant $c$ in Einstein's relativity. However, it is speculated from quantum gravity theories that Lorentz invariance is violated at the Planck scale ($E\sim E_{\rm Pl} = \sqrt{\hbar c^5 / G}\approx 1.22 \times 10^{19} ~ \rm GeV$)~\cite{lrr-2005-5}. A possible consequence is that the speed of light is no longer a constant $c$ but energy dependent. For photon with energy $E\ll E_{\rm Pl}$, the dispersion relation can be written in a theory-independent form as the leading terms in Taylor series
\begin{equation}\label{eq:dispersion relation}
  v(E)=c\left[1-s_n\frac{n+1}{2}\left(\frac{pc}{E_{\mathrm{LV},n}}\right)^n\right],
\end{equation}
where $n=1$ or 2 is usually assumed as the leading power for the correction to the light speed, $s_n=\pm1$ indicates whether the high energy photon travels slower ($s_n=+1$) or faster ($s_n=-1$) than the low energy photon, and $E_{{\mathrm{LV},}n}$ denotes the $n$th-order Lorentz invariance violation~(LV)~scale to be determined.
Such a speed deviation is too small to be detected by most experiments or observations. Amelino-Camelia {\it et al.}~\cite{method1, method2} first suggested to use the data of Gamma Ray Bursts (GRBs) to test this deviation. The high energy (over 10~GeV) of photons from GRB emissions, together with the long distance between the Earth and the GRB source (with redshift up to 8), could produce an observable difference between the arrival times of photons with different energies. Various methods have been applied to search for the light speed variation from GRB data, see, e.g., Refs.~\cite{oldformula,Ellis_app,Rodri,Lamon_grg,DisCan,Abdo_1,Abdo_2,Xiao:2009xe,shaolijing,SMM,zhangshu,Vlasios Vasileiou,Xu_app,Xu_plb,note-added}.
One essential question is that if one could
have a general analysis without any bias to arrive at a reliable
conclusion from experimental data.

A remarkable regularity was noticed in previous studies~\cite{Xu_app,Xu_plb} to look for the light speed variation from high energy photon events of different GRBs. Such regularity was also examined in a recent study~\cite{note-added} to check for {\it in vacuo} dispersion features of gamma ray burst neutrinos and photons.
The aim of this article is to provide a general analysis on such regularity from the data of 25 bright GRBs detected by Fermi Gamma-ray Space Telescope (FGST)~\cite{LAT,GBM}.
Compared to previous intuitive analysis~\cite{Xu_app,Xu_plb} with the help of added straight lines by hand, the present method provides a completed scan over all possibilities in a more clean and impartial way without
any biased choice of straight lines, therefore the conclusion is more reliable and convincing.
The results show that for photons with intrinsic energy (energy at the GRB source) higher than 40~GeV, such regularity exists at a significance of 3-5 $\sigma$ with $E_{\rm LV}=3.6\times 10^{17}~\rm GeV$ determined by the data.

We focus on the $n=1$ case in Eq.~(\ref{eq:dispersion relation}) and redundant subscripts are hereafter omitted. For two photons with energies $E_{\rm high}$ and $E_{\rm low}$ respectively, the dispersion relation in Eq.~(\ref{eq:dispersion relation}) would lead to a difference between their observation times as~\cite{oldformula,newformula}
\begin{equation}\label{eq:t_LV}
  \Delta t_{\mathrm{LV}}= (1+z) \frac{\kappa}{E_{\mathrm{LV}}},
\end{equation}
where
\begin{equation}\label{eq:kappa}
  \kappa=s\frac{E_{\mathrm{high}}-E_{\mathrm{low}}}{H_0} \frac{1}{(1+z)} \int_0^z\frac{(1+z')\mathrm{d}z'}
{\sqrt{\Omega_{\mathrm{m}}(1+z')^3+\Omega_{\Lambda}}}
\end{equation}
is the Lorentz violation factor, $z$ is the redshift of the GRB source, $H_0=\mathrm{67.3\pm 1.2 ~km s^{-1} Mpc^{-1} }$ is the Hubble expansion rate, and $[\Omega_m, \Omega_{\Lambda}]=[\mathrm{0.315^{+0.016}_{-0.017}}, \mathrm{0.685^{+0.017}_{-0.016}}]$ are cosmological constants~\cite{pgb}.

By taking $\Delta t_{\mathrm{LV}}$ into consideration, the intrinsic time lag at the GRB source between two photons is
\begin{equation}\label{eq:intrinsic_lag}
  \Delta t_{\rm in}=\frac{\Delta t_{\rm obs}}{1+z}-\frac{\kappa}{E_{\rm LV}},
\end{equation}
where $\Delta t_{\rm obs}$ is the observed time lag between these two photons.
The intrinsic time lag $\Delta t_{\rm in}$ is a function of $E_{\rm LV}$
due to the fact that the time delay caused by the light speed
variation should be subtracted from the observed time lag.
In our analysis, the energy threshold of high energy photons is set as 1~GeV for intrinsic energy, i.e., $E_{\rm high}(1+z) \geqslant 1~\rm GeV$. These photons are detected by the Large Area Telescope (LAT)~\cite{LAT}~onboard FGST. On the other hand, $E_{\rm low}$ is lower than $260~\rm keV$~\cite{GBM}~and thus can be omitted in Eq.~(\ref{eq:kappa}). Obviously, if we fix the low energy photon, the distribution of $\Delta t_{\rm in}$ would be the light curve of high energy emission at the GRB source. For a certain GRB, it is in fact arbitrary to choose such a low energy photon (different choices only lead to a shift of the zero point). But since we are going to use data from different GRBs, a unified standard is required. Here, we follow Refs.~\cite{Xu_app,Xu_plb} where the low energy photon is chosen as the first main peak of Gamma-ray Burst Monitor (GBM)~\cite{GBM}~light curve, with photon energies ranging between $8 \sim 260$~keV.
The
criteria for the low energy photon peak time $t_{\mathrm{peak}}$ can be performed
under difference conditions such as in the observer reference
frame and in the source reference frame, with also different
energy bands and time bins, and these different choices can lead
to a uncertainty of around 1 second for each $t_{\mathrm{peak}}$. Such
uncertainties have little influences on the analysis in this
work.
We thus can adopt the ``intrinsic time" $t=\Delta t_{\rm in}$ as the high energy photon emitting time at the source for our simultaneous analysis of different GRBs. The intrinsic properties of different GRBs are not the same from a strict sense. We assume that GRBs have similarities in some respects, such as the correlation between the high energy event and a remarkable low energy event in the GRB source reference frame, and thus different GRBs could be analyzed simultaneously to reveal possible regularities behind the GRB data.

The method we apply here
can trace back to the concept of ``information entropy''~\cite{information_entropy} ~and is similar to the ``dispersion cancellation'' (DisCan) method~\cite{DisCan}, the ``minimal dispersion'' method~\cite{minimum_disp}, the ``energy cost function'' method~\cite{energy_cost_fun}, and the ``sharpness-maximization'' method (SMM)~\cite{SMM}. The essence of this method is based on the expectation that the distribution of emitting time $t$, i.e., the light curve of GRBs, should be ``sharp'' at the source and the dispersion relation~(\mbox{\ref{eq:dispersion relation}})~tends to smear such a sharp feature. Therefore, a trial of true $E_{\rm LV}$ in $\Delta t_{\mathrm{LV}}$ of Eq.~(\mbox{\ref{eq:t_LV}}) would recover the maximal sharpness whereas a trial of wrong one would not be so effective. To be specific, with each trial $E_{\rm LV}$, we can obtain the emitting times of all photons in a data set with Eq.~(\ref{eq:intrinsic_lag}). The sharpness of the distribution would reach its maximum when the trial $E_{\rm LV}$ equals to its true-value.

The next step is to quantify the sharpness of the distribution. We adopt a function $\mathcal{S}$ defined as
\begin{equation}\label{eq:S}
  \mathcal{S}(E_{\rm LV})=\sum_{i=1}^{N-\rho}\log \left(\frac{\rho}{t_{i+\rho}-t_{i}}\right),
\end{equation}
where $t_{i}$ is the emitting time of the $i$-th photon in the data set (sorted in ascending order). It should be noted that in~\cite{SMM}, $t'$ (see Eq.~(8) therein) is measured on the Earth (detectors), while in our present study, $t$ is the intrinsic time at the GRB source to accommodate different GRBs.

In Eq.~(\ref{eq:S}), $\rho$ is a pre-set integer parameter. For a too small $\rho$, $\mathcal{S}$ would fluctuate significantly with $E_{\rm LV}$ and the accuracy of the present method would be undermined. For a too large $\rho$, the $\mathcal{S}$-$E_{\rm LV}$ curve would be too smooth and the peak of $\mathcal{S}$ might be flattened. In our study, $\rho$ is set as 5, by considering the size of our data sets. In fact, the choice of $\rho$ has little influence on the main conclusions of our work as long as it is reasonable
(see 
\mbox{\ref{app2}}~for more information). A test of the validity of our method is performed on some artificially produced photon events, see \mbox{\ref{app3}}.

Our analysis exhaust all LAT photons that have intrinsic energy higher than 1~GeV, have known redshift and are within a 90 second time window~\cite{LATdata}. The redshifts of these GRBs are determined by
the light spectrum with high precision. 524 photons from 25 GRBs are included (see Tab.I
in 
\ref{app1}). From them, four different data sets are constructed with photons
\begin{enumerate}
  \item  with intrinsic energies between $1\sim 10~\rm GeV$, with 481 photons in total;
  \item  with intrinsic energies above 10~GeV, with 43 photonos in total;
  \item  with intrinsic energies above 20~GeV, with 20 photons in total;
  \item  with intrinsic energies above 40~GeV, with 12 photons in total.
\end{enumerate}
There are only 2 photons with intrinsic energies between $20\sim 30~\rm GeV$ and the data set ``over 30~GeV" has similar property to that of data set~\uppercase\expandafter{\romannumeral3}. It is thus omitted for simplicity.

\begin{figure}[tbp]
  \centering
  \includegraphics[width=0.48\textwidth]{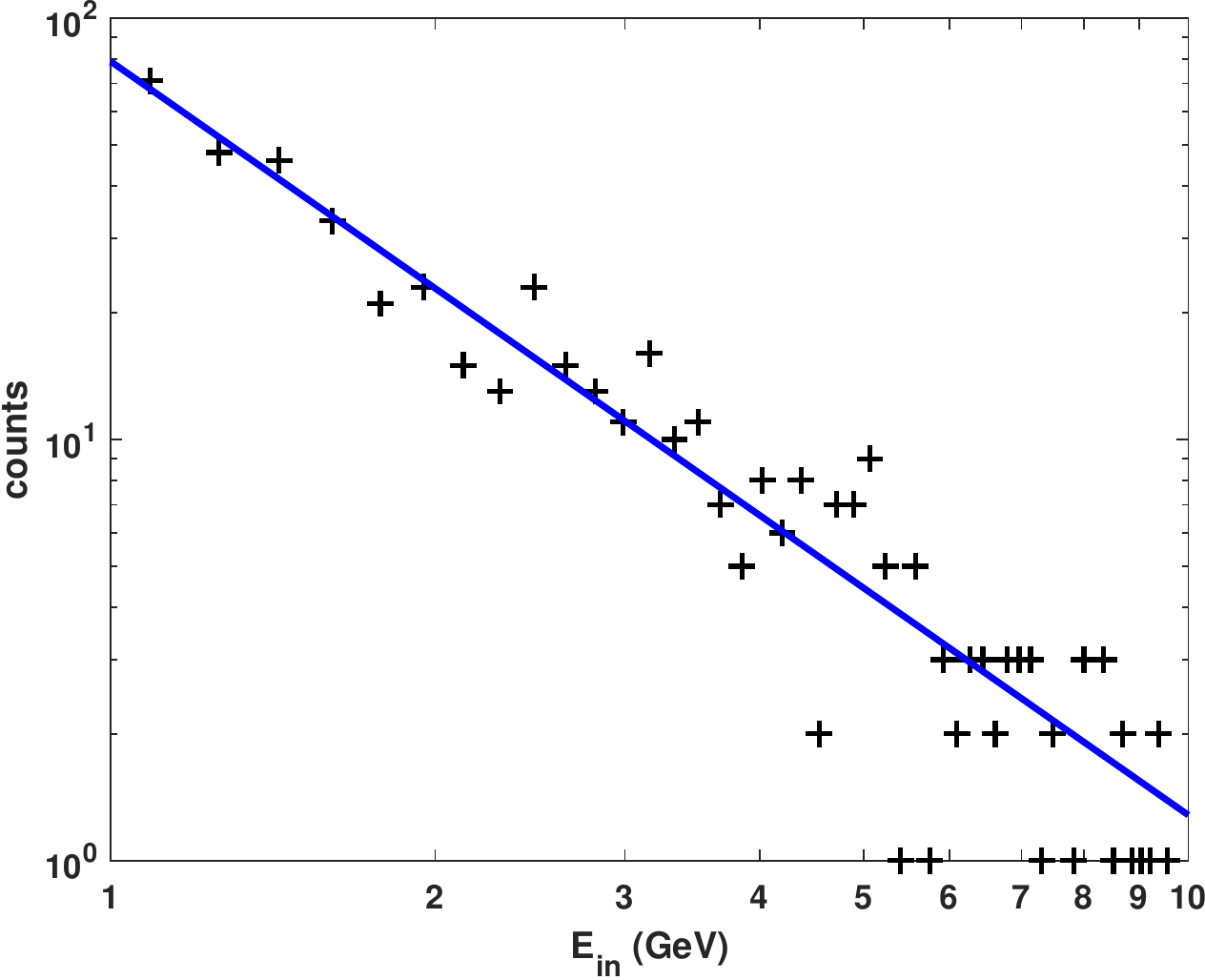}
  \caption{The intrinsic energy spectrum of photons in data set~\uppercase\expandafter{\romannumeral1}~(unnormalized). This spectrum can be fitted by a power function $\mathrm{d}N/ \mathrm{d} E \propto E^{-1.8}$ (the blue curve).}\label{fig:E_in_fit}
\end{figure}

In order to understand better the results of our analysis and to test the robustness of our method, we analyze some random data, in comparison with the observed data. The random data sets are produced as follows:
\begin{enumerate}
  \item the size of the random data set is the same as that of the observed data set;
  \item the redshift of the observed data are simply copied onto the random data;
  \item the arrival time of observed data are randomly permutated and then used as that of random data. In other words, the corresponding relationship between redshift and arrival time is randomly exchanged;
  \item for intrinsic energy, the spectrum of observed photons in data set~\uppercase\expandafter{\romannumeral1}~is fitted with a power function (see Fig.~\ref{fig:E_in_fit}). The result is
      \begin{equation}\label{eq:E_in_fit}
         \frac{\mathrm{d} N}{\mathrm{d} E} \propto E^{-\alpha}, \qquad \qquad \alpha=1.8.
      \end{equation}
      Then, the spectrum is extrapolated into higher ranges (for data sets ~\uppercase\expandafter{\romannumeral2}~$\sim$~\uppercase\expandafter{\romannumeral5}). The energies of random data set are produced to satisfy this distribution and have the same range with the observed data set.
\end{enumerate}
Every time a random data set is produced as above, we get a $\mathcal{S}$-$E_{\rm LV}$ curve from it.

\begin{figure}
  \begin{center}
  \subfloat[]{
  \label{fig:1a}
    \includegraphics[width=0.48\textwidth]{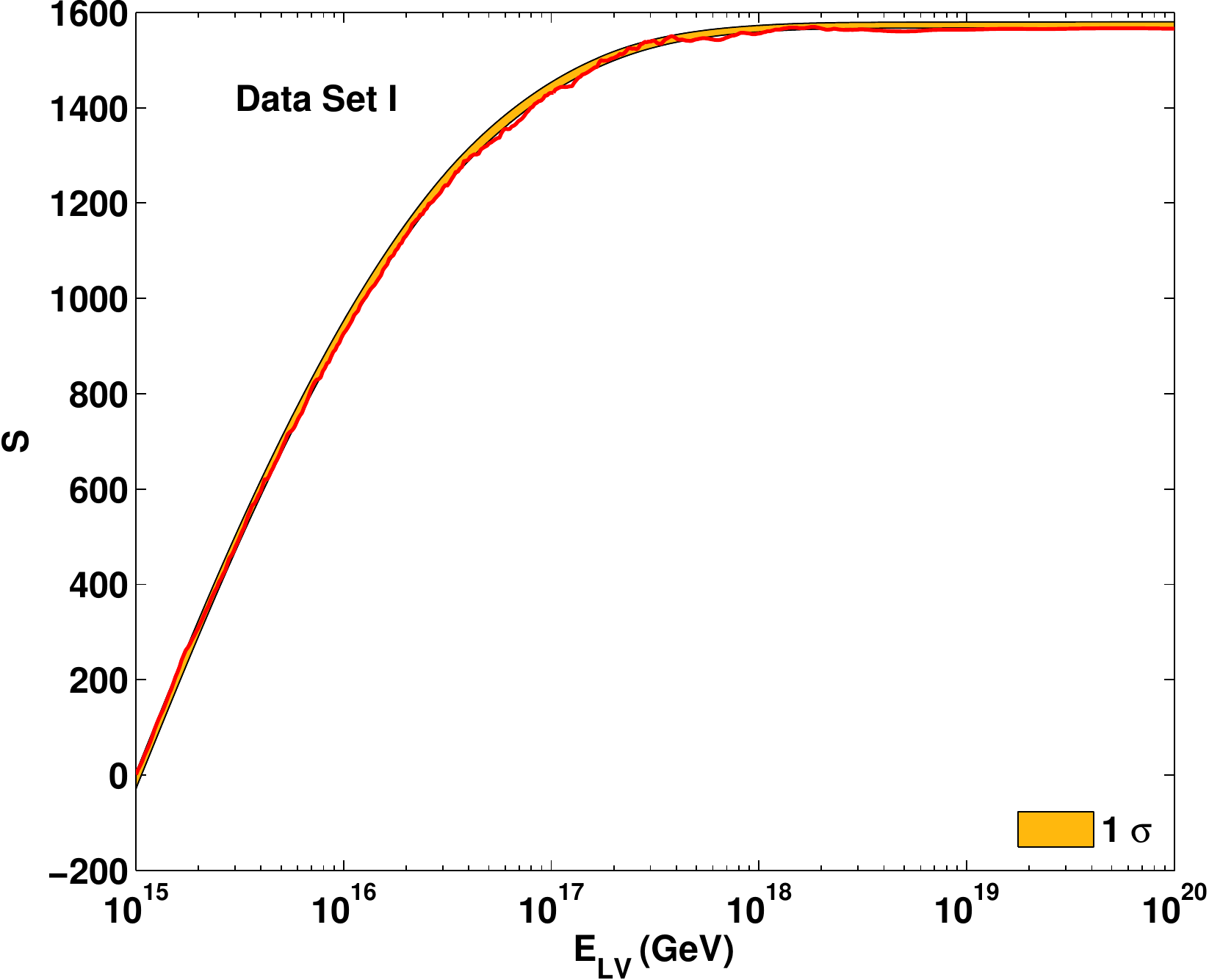}
    }
  \subfloat[]{
  \label{fig:1b}
    \includegraphics[width=0.48\textwidth]{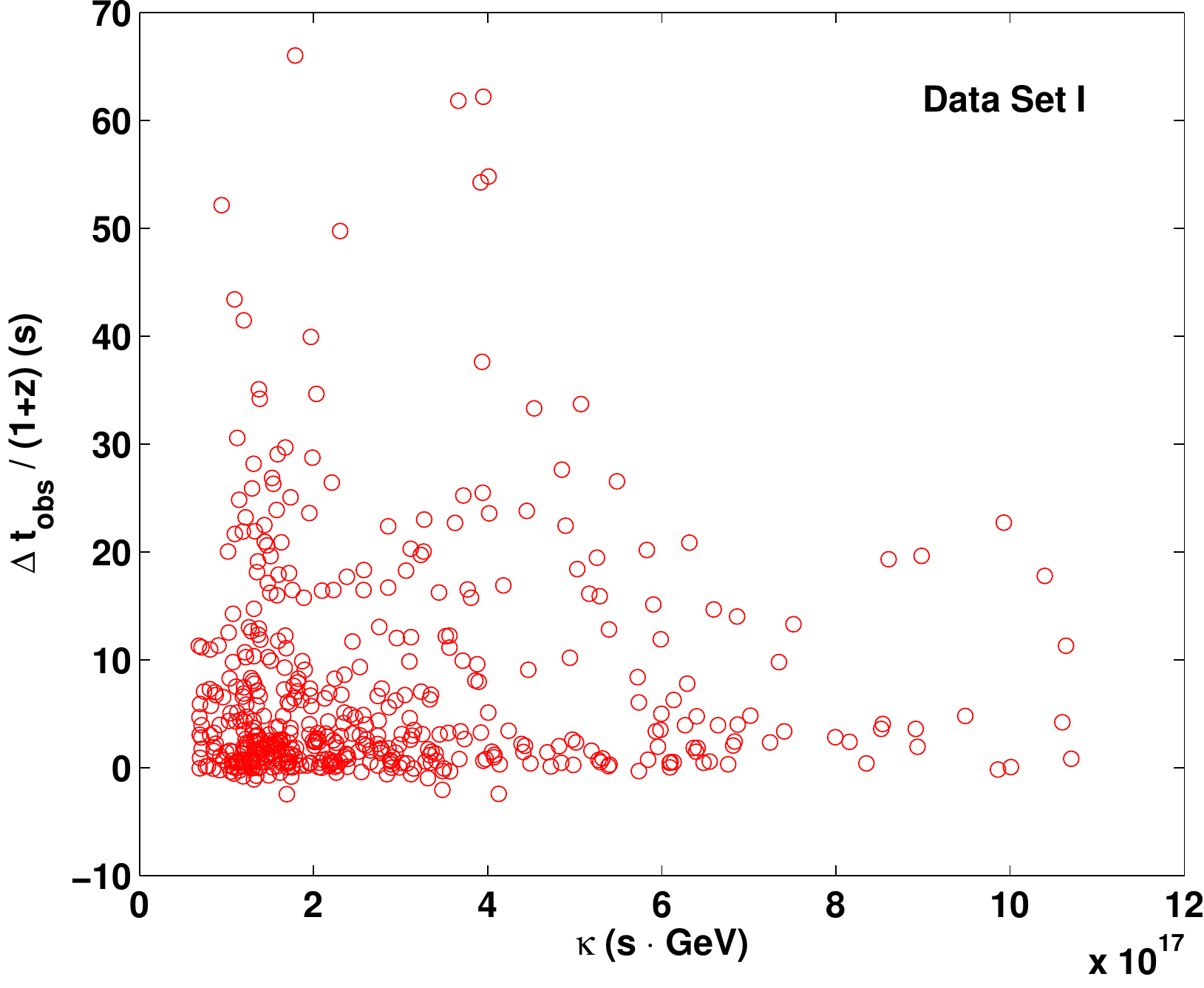}
  }
  \end{center}
  \caption{In the left panel, the red curve is the $\mathcal{S}$-$E_{\rm LV}$  curve for data set~\uppercase\expandafter{\romannumeral1} and two thin curves enclose the $1\,\sigma$ region for random data. The right panel is the~$\Delta t_{\rm obs}/(1+z)$-$\kappa$~plot for all photons in data set~\uppercase\expandafter{\romannumeral1}.}\label{fig:data_set_1}
\end{figure}

\begin{figure}
  \centering
  \subfloat[]{
  \label{fig:2a}
    \includegraphics[width=0.48\textwidth]{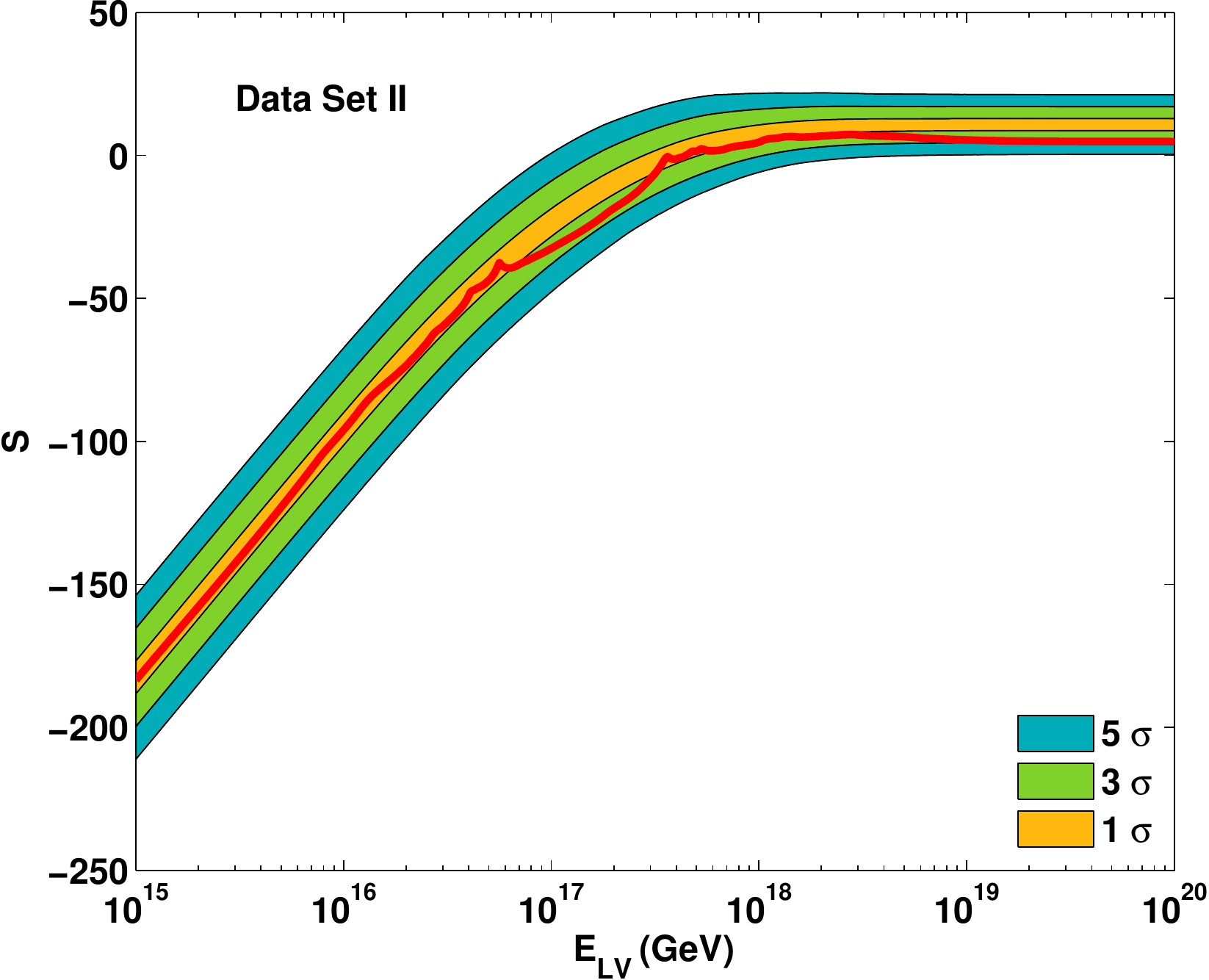}
    }
    \subfloat[]{
  \label{fig:2b}
    \includegraphics[width=0.48\textwidth]{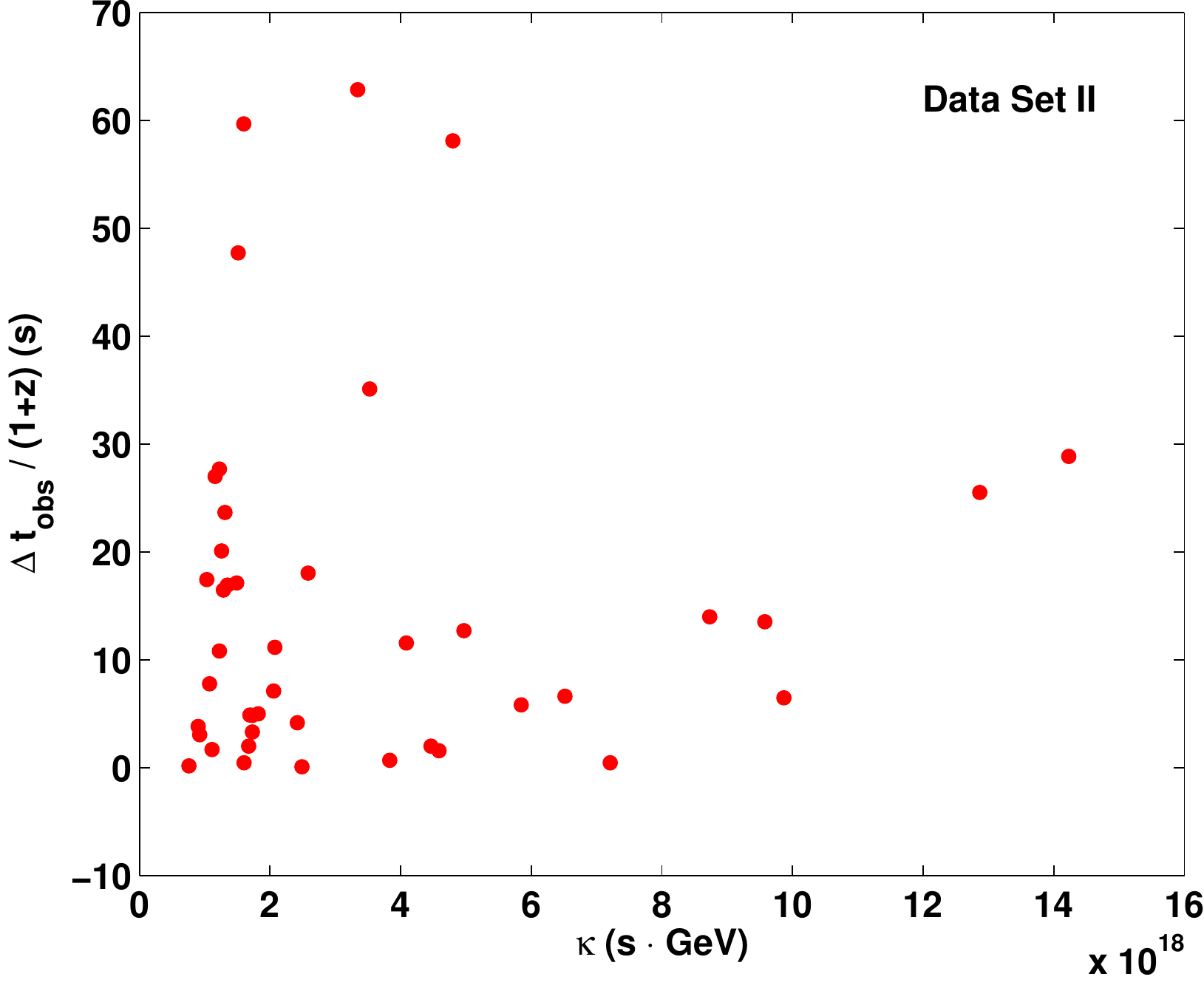}
    }
    \\
    \subfloat[]{
  \label{fig:2c}
    \includegraphics[width=0.48\textwidth]{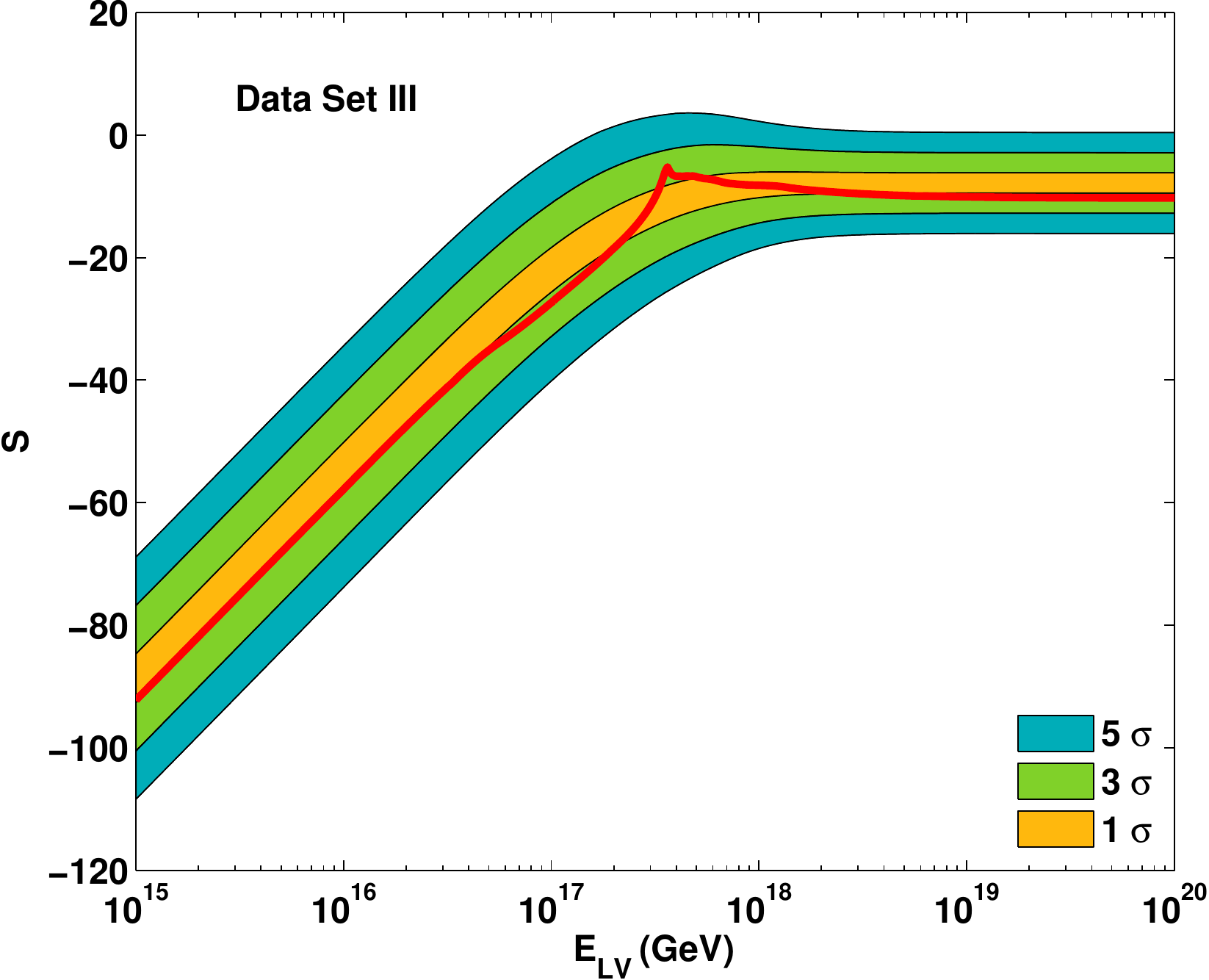}
    }
    \subfloat[]{
  \label{fig:2d}
    \includegraphics[width=0.48\textwidth]{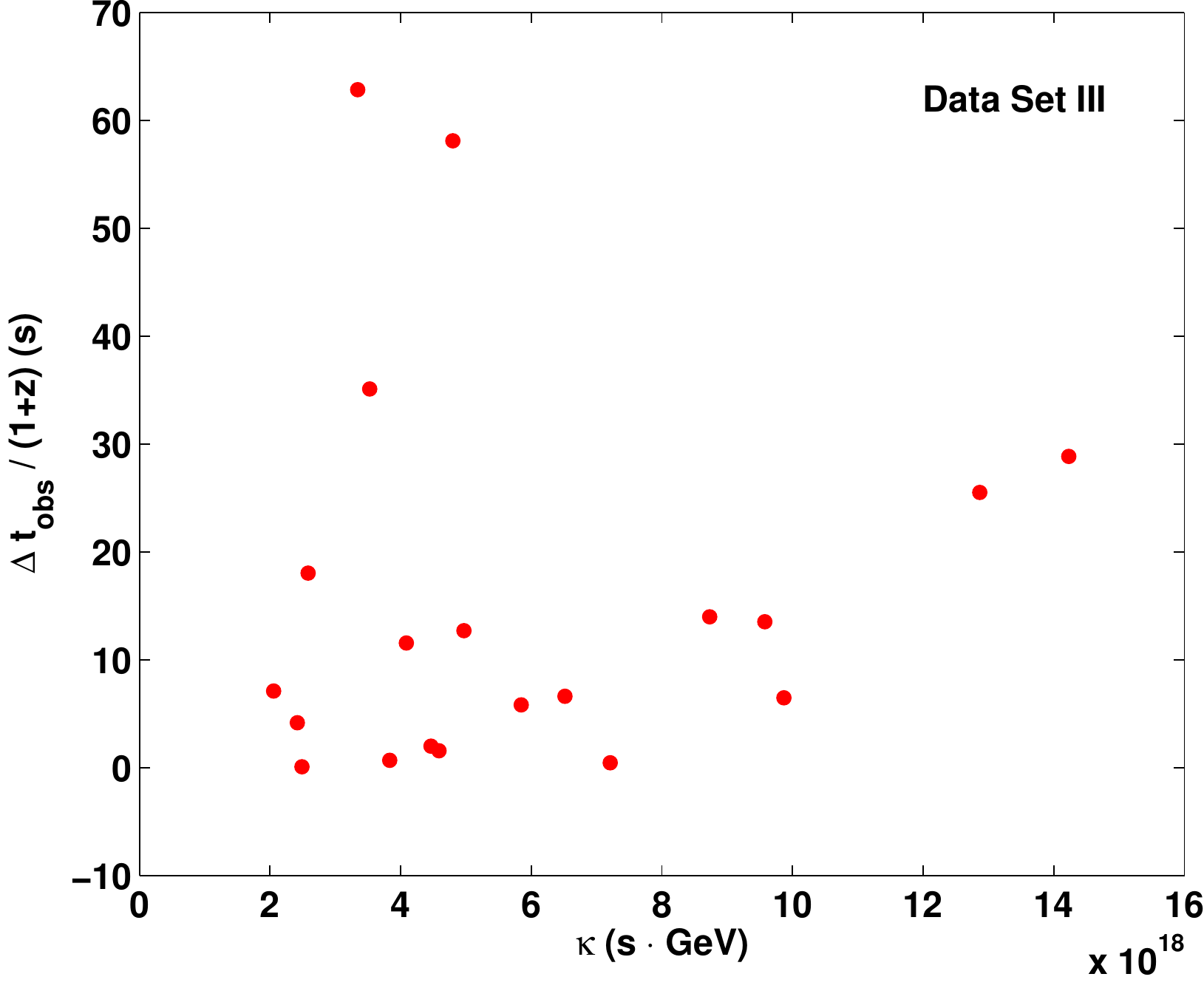}
    }
  \\
  \subfloat[]{
  \label{fig:2e}
    \includegraphics[width=0.48\textwidth]{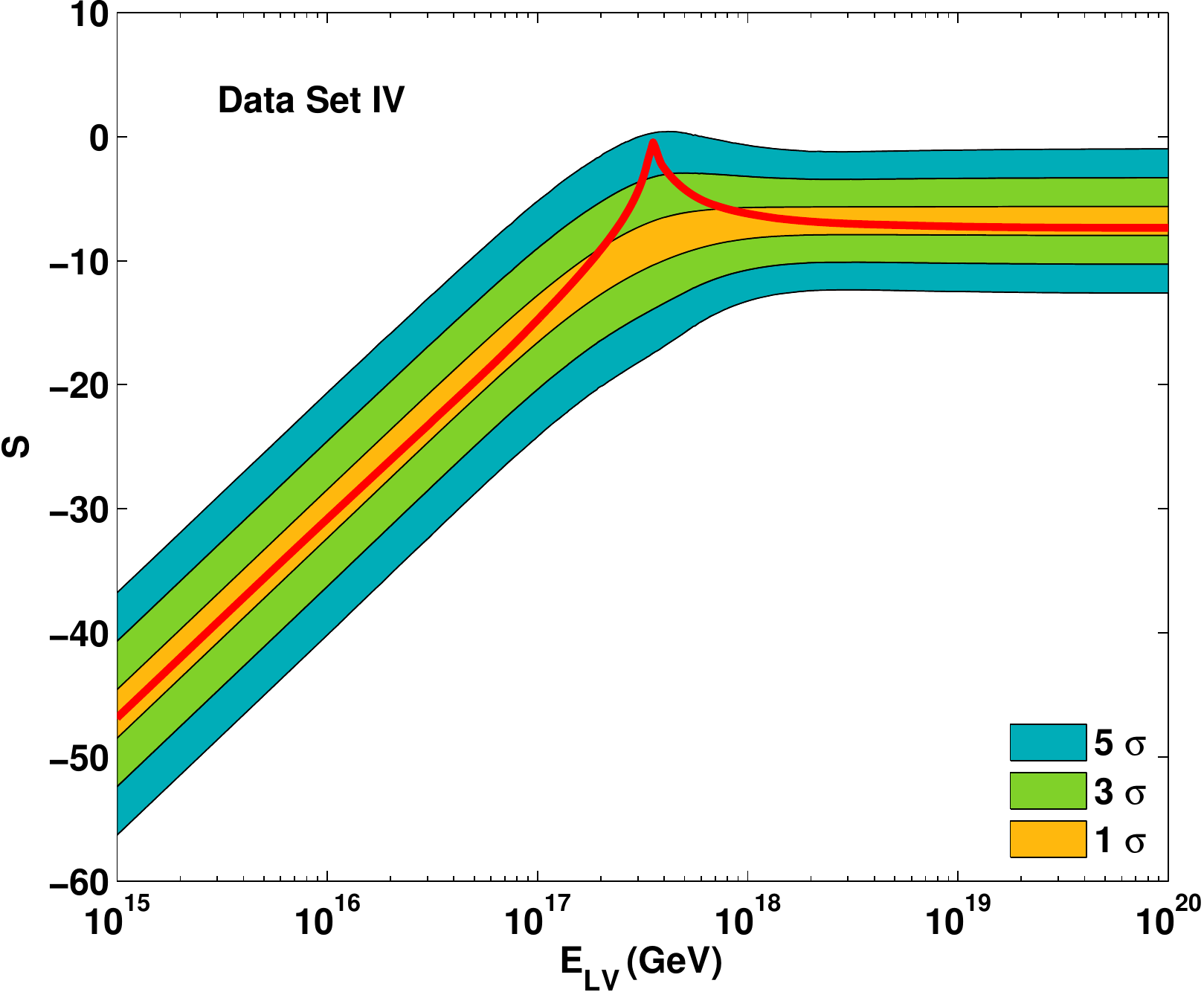}
    }
    \subfloat[]{
  \label{fig:2f}
    \includegraphics[width=0.48\textwidth]{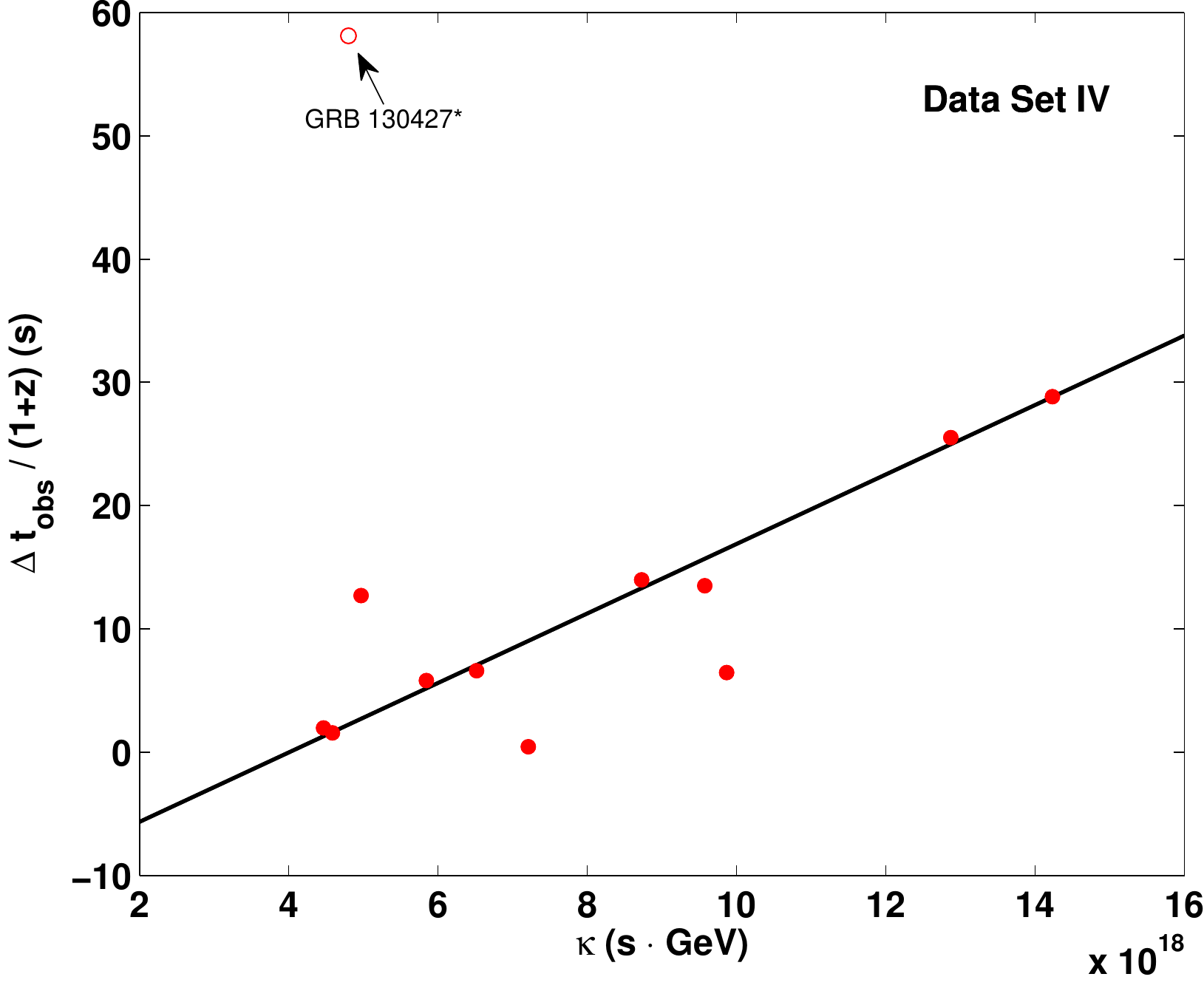}
    }
    \caption{In the left panels, the red curves are the $\mathcal{S}$-$E_{\rm LV}$  curves of observed data while yellow, green and blue areas are the $1\,\sigma$, $3\,\sigma$ and $5\,\sigma$ regions for random data respectively. The right panels are the~$\Delta t_{\rm obs}/(1+z)$-$\kappa$~plots for photons analyzed in the left panels. Upper and lower panels correspond to data sets~\uppercase\expandafter{\romannumeral2}~$\sim$~\uppercase\expandafter{\romannumeral4} respectively. In Fig.~\ref{fig:2f}, the slope of the black straight line is $1/E_{\rm LV}$, where $E_{\rm LV}$ is determined by the $\mathcal{S}$-$E_{\rm LV}$  curve in Fig.~\ref{fig:2e}.}\label{fig:data_set_2_to_4}
\end{figure}

The results of our analysis are shown in the left panels of Figs.~\ref{fig:data_set_1},~\ref{fig:data_set_2_to_4}. In all cases, $s$ in Eq.~(\ref{eq:kappa})~is $+1$. The red curves are $\mathcal{S}$-$E_{\rm LV}$ curves for observed data in data sets~\uppercase\expandafter{\romannumeral1}~$\sim$~\uppercase\expandafter{\romannumeral4}~respectively. We can see that for data sets~\uppercase\expandafter{\romannumeral1}~and~\uppercase\expandafter{\romannumeral2}, the $\mathcal{S}$-$E_{\rm LV}$ curves have no obvious peaks, except for some small bulges. However, a peak emerges in the $\mathcal{S}$-$E_{\rm LV}$ curve of data set~\uppercase\expandafter{\romannumeral3}~and becomes more significant for data set~\uppercase\expandafter{\romannumeral4}.
A completed scan over the whole $E_{\rm LV}$ range (i.e., with trial $E_{\rm LV}=-10^{40} \to 10^{40}$~GeV) finds no other peaks except the above one.

In order to clarify the confidence level that our results can be trusted, random data are used as a comparison. For each observed data set, we produce $10^{5}$ random data sets. Therefore, for each $E_{\rm LV}$, there are $10^5 \,\,\mathcal{S}$ generated from random data. Since random data sets are different from each other, these $\mathcal{S}$ are not identical, but approximately follow Gaussian distribution. The expectation $\mu$ and standard deviation $\sigma$ can easily be obtained. Next, we go through all trial $E_{\rm LV}$ and draw the $1\,\sigma$ region $(\mu-\sigma,~\mu+\sigma)$ for each $E_{\rm LV}$. This leads to a $1\,\sigma$ region for the $\mathcal{S}$-$E_{\rm LV}$ curve of random data (see the yellow areas Figs.~\ref{fig:data_set_1},~\ref{fig:data_set_2_to_4}). Similarly, we can get the $3\,\sigma$ and $5\,\sigma$ regions for random data (see the green and blue areas in Figs.~\ref{fig:data_set_2_to_4} respectively).
\begin{figure}
  \centering
\subfloat[]{
  \label{fig:3a}
    \includegraphics[width=0.48\textwidth]{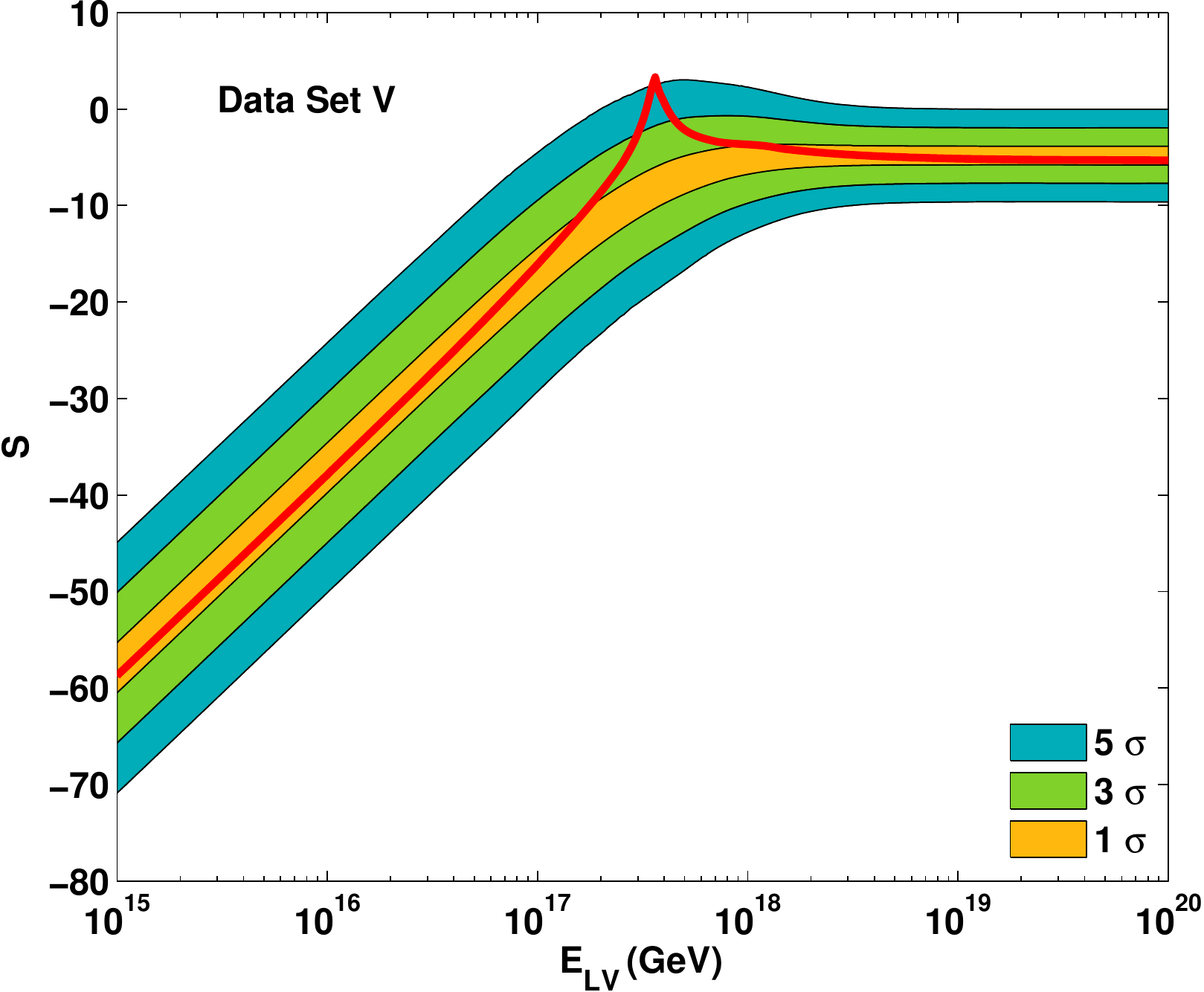}
    }
    \subfloat[]{
  \label{fig:3b}
    \includegraphics[width=0.48\textwidth]{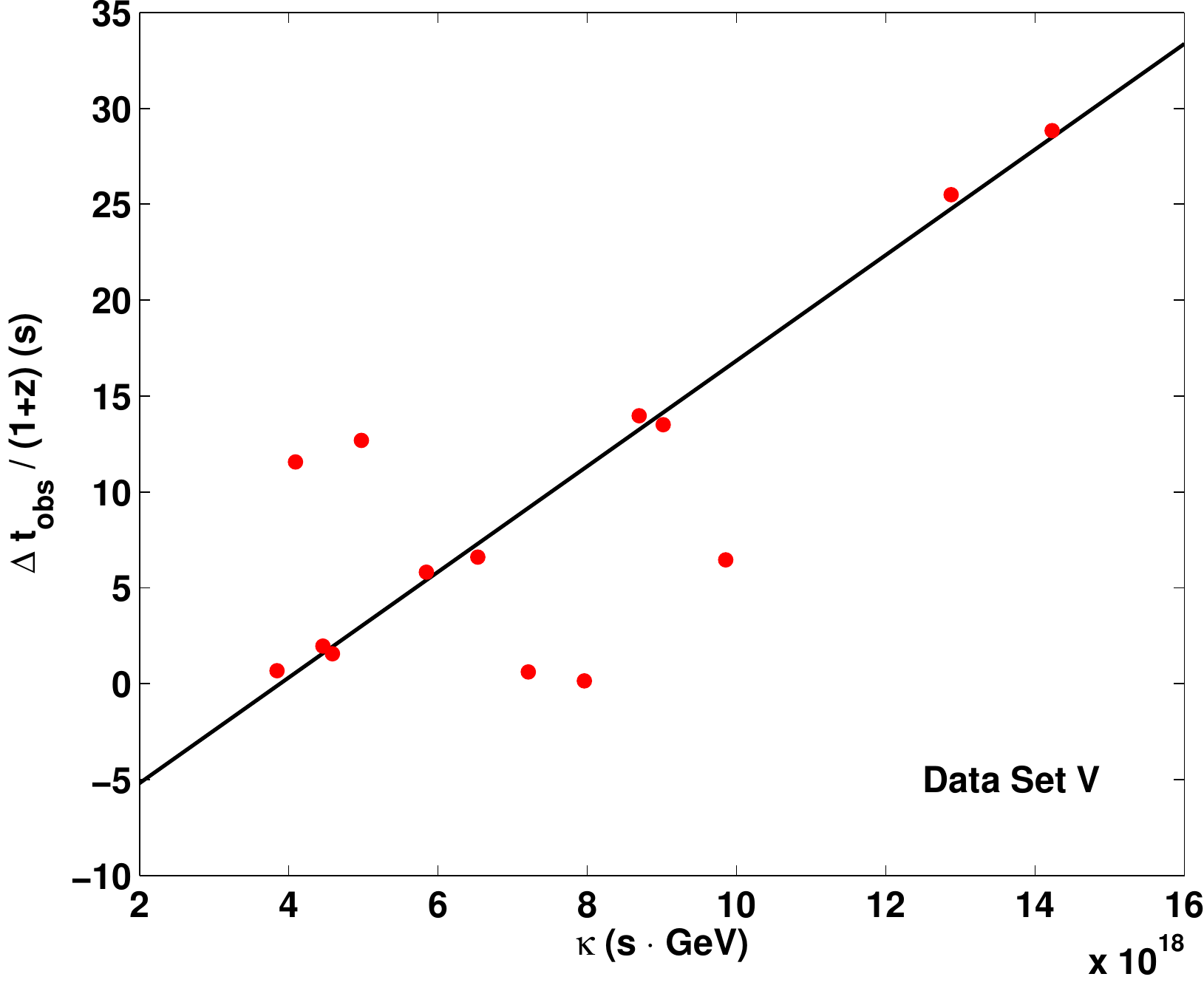}
    }
  \caption{In the left panel, the red curves are the $\mathcal{S}$-$E_{\rm LV}$  curves of observed data in data set~\uppercase\expandafter{\romannumeral5}~(published in Ref.~\cite{Xu_plb}) while yellow, green and blue areas are the $1\,\sigma$, $3\,\sigma$ and $5\,\sigma$ regions for random data respectively. The right panel is the~$\Delta t_{\rm obs}/(1+z)$-$\kappa$~plot~(published in Ref.~\cite{Xu_plb}) for photons analyzed in the left panel. The slope of the red straight line is $1/E_{\rm LV}$, where $E_{\rm LV}$ is determined by the $\mathcal{S}$-$E_{\rm LV}$  curve in Fig.~\ref{fig:3a}.}\label{fig:data_set_5}
\end{figure}

\begin{figure}
  \centering
  \includegraphics[width=0.48\textwidth]{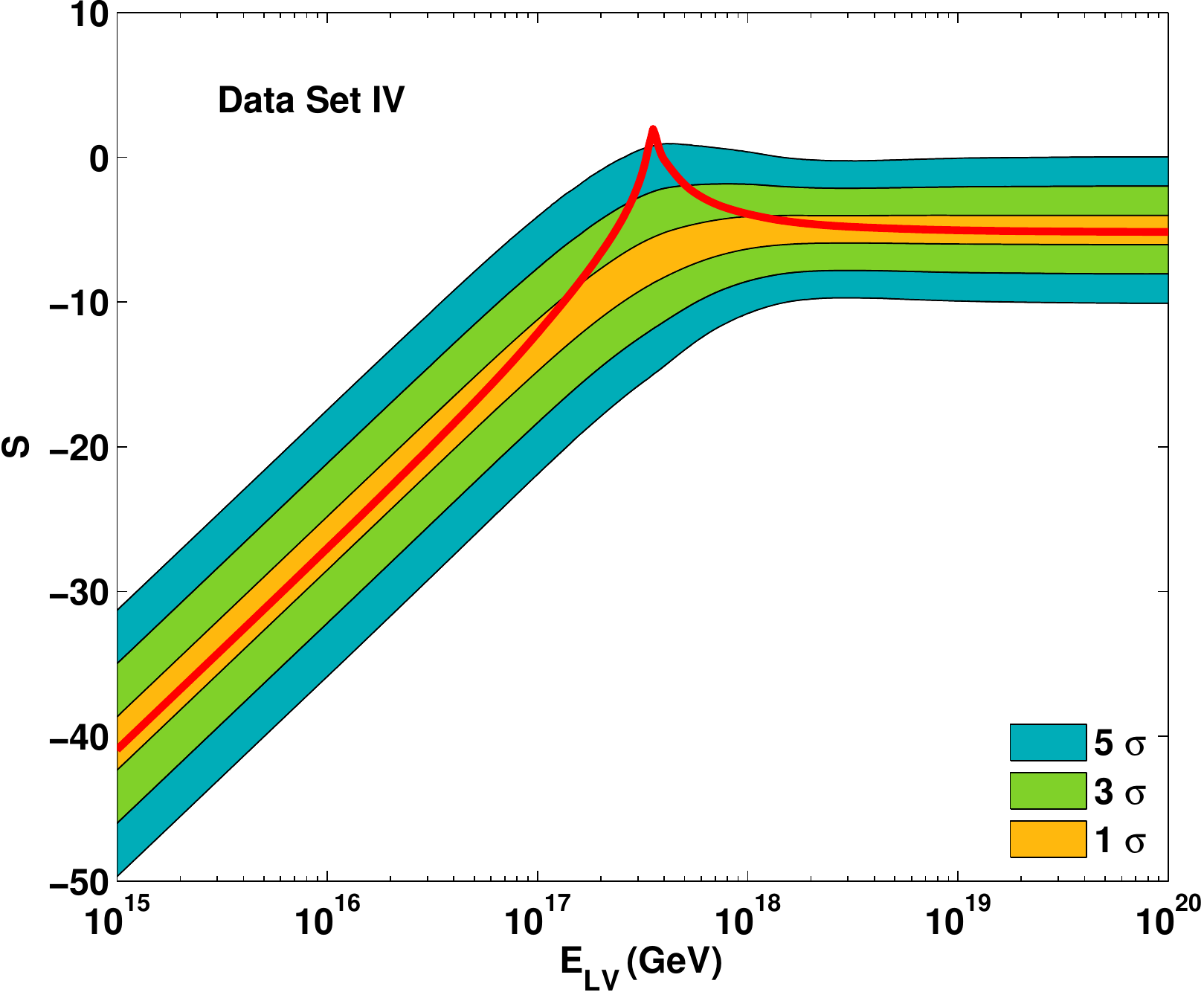}
  \caption{The $\mathcal{S}$-$E_{\rm LV}$  curve for data set~\uppercase\expandafter{\romannumeral4} without the photon GRB~$130427^{\star}$.}\label{fig:exclude_one_photon}
\end{figure}

If the $\mathcal{S}$-$E_{\rm LV}$  curve of a observed data set deviates significantly from that of a random data set, it can be concluded that there are regularities behind the observed data. In Fig.~\ref{fig:data_set_1}, nearly whole $\mathcal{S}$-$E_{\rm LV}$  curve for the observed data is inside the $1\,\sigma$ region for random data. Therefore, we cannot exclude the possibility that the observed data set~\uppercase\expandafter{\romannumeral1} is randomly distributed. For data sets~\uppercase\expandafter{\romannumeral2}~and~\uppercase\expandafter{\romannumeral3}, some parts of their $\mathcal{S}$-$E_{\rm LV}$  curves are outside the $1\,\sigma$ region for random data, but whole of these curves are still inside the $3\,\sigma$ region~(see Figs.~\ref{fig:2a},~\ref{fig:2c}). However, the peak of the $\mathcal{S}$-$E_{\rm LV}$  curve for data set~\uppercase\expandafter{\romannumeral4}~is well outside the $3\,\sigma$ region for random data and nearly goes beyond the $5\,\sigma$ region~(see Fig.~\ref{fig:2e}).
We can conclude that a finite $E_{\rm LV}$ is favored 
at a significance of $3\,\sigma$ ($5\,\sigma$).

It seems that we cannot distinguish the behavior of photons with intrinsic energies lower than 10~GeV from that of randomly produced photons. But with the increase in energy threshold, regularities emerge and gradually become more significant.

For data set~\uppercase\expandafter{\romannumeral4}, the peak of its $\mathcal{S}$-$E_{\rm LV}$  curve is at $E_{\rm LV}=3.55\times 10^{17}~\rm GeV$. This coincides well with the result of Refs.~\cite{Xu_app,Xu_plb}, where $E_{\rm LV}$ is determined to be $3.60\times 10^{17}~\rm GeV$, but obtained in a different way. There, Eq.~(\ref{eq:intrinsic_lag})~is expressed as
\begin{equation}\label{eq:linear_relation}
    \frac{\Delta t_{\rm obs}}{1+z}=\frac{\kappa}{E_{\rm LV}}+\Delta t_{\rm in}
\end{equation}
and all photons are drawn on the~$\Delta t_{\rm obs}/(1+z)$-$\kappa$~plot to determine $E_{\rm LV}$ as the reciprocal of the slope of the mainline (see Fig.~2 in~\cite{Xu_plb}). It is worth noting that this method and our present general method are complementary to each other. On the one hand, there might be some bias when drawing straight lines in the~$\Delta t_{\rm obs}/(1+z)$-$\kappa$~plots~(see the discussions in~\cite{Xu_app,Xu_plb}). But our general method makes up for this deficiency since it is impartial to all possible values of $E_{\rm LV}$. Let us use our present method to re-analyze data in Ref.~\cite{Xu_plb}, named as data set~\uppercase\expandafter{\romannumeral5}. This data set contains 14 photons that have observed energy higher than 10~GeV, have known redshift and are within a 90 second time window. The $\mathcal{S}$-$E_{\rm LV}$  curve for data set~\uppercase\expandafter{\romannumeral5} is shown in Fig.~\ref{fig:3a}. A distinct peak at $E_{\rm LV}=3.63\times 10^{17}~\rm GeV$ goes beyond the $5\,\sigma$ region for random data. Therefore, the result in Ref.~\cite{Xu_plb} is in fact reliable at a significance of $5\,\sigma$.

On the other hand, the peaks of $\mathcal{S}$-$E_{\rm LV}$  curves can be illustrated in a more intuitive way by straight lines in~$\Delta t_{\rm obs}/(1+z)$-$\kappa$~plots. We draw the $\Delta t_{\rm obs}/(1+z)$-$\kappa$~plots for data sets~\uppercase\expandafter{\romannumeral1}~$\sim$~\uppercase\expandafter{\romannumeral4} respectively, as shown in the right panels of Figs.~\ref{fig:data_set_1},~\ref{fig:data_set_2_to_4}. In Fig.~\ref{fig:1b}, the distribution of the points are rather messy and thus no straight lines can be drawn. This is consistent with the fact that no peaks appear in the $\mathcal{S}$-$E_{\rm LV}$  curve of data set~\uppercase\expandafter{\romannumeral1}. However, with the increase in the energy threshold, the distribution of points in $\Delta t_{\rm obs}/(1+z)$-$\kappa$~plots gradually becomes regular. Finally, a straight line can be drawn in the plot for data set~\uppercase\expandafter{\romannumeral4} (Fig.~\ref{fig:2f}). This is also consistent with the distinct peak of $\mathcal{S}$-$E_{\rm LV}$  curve for data set~\uppercase\expandafter{\romannumeral4}.
It is interesting that in Fig.~\ref{fig:2f}, a point (GRB~130427$^{\star}$) is far away from other points. It seems reasonable to assume that this photon has different properties from other photons and we can delete it from data set~\uppercase\expandafter{\romannumeral4}.
In fact, the same data set without the GRB~130427$^{\star}$ event can be naturally constructed with an alternative selection criteria
of a new intrinsic time window, as shown in a recent study~\cite{note-added}.
The result of present analysis for the new data set~\uppercase\expandafter{\romannumeral4} is shown in Fig.~\ref{fig:exclude_one_photon}. The peak of $\mathcal{S}$-$E_{\rm LV}$  is well outside the $5\,\sigma$ region.

As mentioned earlier, the result of the present work is consistent with that of Refs.~\mbox{\cite{Xu_app,Xu_plb}}. In fact, the idea of simultaneous analysis of multiple GRBs traces back to Ref.~\mbox{\cite{Ellis_app}}. In Refs.~\mbox{\cite{shaolijing,zhangshu}}, the $\Delta t_{\rm obs}/(1+z)$-$\kappa$~plot method was applied to analyze the Fermi GRB data of high energy photons available at that time.
In Ref.~\mbox{\cite{Xu_app}}, the low energy photon arrival time is changed from the trigger time to the first main pulse of low energy photons. With this more natural choice of $t_{\rm low}$, a stronger regularity emerges with 8 out of 13 photon events falling on a same line in the $\Delta t_{\mathrm{obs}}/(1+z)$-$K_{1}$ plot, so a linear form ($n=1$)
light speed variation was suggested at a scale of $E_{\rm LV}=(3.60 \pm 0.26) \times 10^{17}~ \rm GeV$. In Ref.~\mbox{\cite{Xu_plb}}, an additional event from GRB~160509 was found to fall exactly on the same mainline above. In fact, any intrinsic property of GRBs should be revealed by careful
analysis of data rather than any \mbox{\emph{ad hoc}} assumption, and any model prediction should be also tested by data. The predictions in Refs.~\mbox{\cite{shaolijing,zhangshu,Xu_app}} have received support by the newly observed energetic photon event of GRB~160509A in Ref.~\mbox{\cite{Xu_plb}}. A recent study~\cite{note-added} also confirmed the calculations in Refs.~\cite{Xu_app,Xu_plb}.
In our present paper, a new method is applied to re-analyze the data. Such method can avoid the bias of drawing a mainline in the $\Delta t_{\rm obs}/(1+z)$-$\kappa$~plot to determine $E_{\rm LV}$ in previous works. With a complete scan of all trial $E_{\rm LV}$, we arrive at a finite $E_{\rm LV}$ which is compatible with previous results in Refs.~\mbox{\cite{shaolijing,zhangshu,Xu_app,Xu_plb}}. The most important aspect of the new work is to clarify the confidence level of previous results.

As a comparison with Ref.~\mbox{\cite{SMM}}, we actually extend the analysis of data from a single GRB to multi-GRBs. In Ref.~\mbox{\cite{SMM}} data of only one GRB are analyzed at a time, whereas we include all high energy events of photons from multiple GRBs. If we only analyse data from one single GRB, we would be also not possible to find any peak in the $E_{\rm LV}$ scanning. Therefore the reason for the difference between our work and Ref.~\mbox{\cite{SMM}} is due to the simultaneous analysis of all high energy photon events in the Fermi GRB data with known redshifts.
In fact, the difference between our work and the Fermi analysis in Ref.~\mbox{\cite{Abdo_2}}
has been discussed in detail in Ref.~\mbox{\cite{zhangshu}}. Ref.~\mbox{\cite{Abdo_2}}
is based on the assumption that the high photon events are not emitted earlier than the low-energy photons at the GRB source, whereas we analyse the high energy photon events from all Fermi GRB data without any biased assumption.
It is pointed in Ref.~\mbox{\cite{zhangshu}} that some GRBs including short GRBs are found to have high energy ($>$0.5~GeV) photons with minus intrinsic time lag.
This seems to be compatible with our results that some high energy photons might be emitted earlier at the source.
Of course, more data are still needed to test different predictions.
In fact, the predictions in Refs.~\mbox{\cite{shaolijing,zhangshu,Xu_app}} have been clearly supported by the newly observed energetic photon event of GRB~160509A in Ref.~\mbox{\cite{Xu_plb}}. As has been discussed in Ref.~\cite{note-added}, the regularity revealed in this manuscript could be an effect of light speed variation in cosmological space, it is also possible due to astrophysical reasons to be explored.

In conclusion, we use a general method to analyze the data of 25 bright GRBs detected by FGST. The results suggest that for photons with energy higher than 40~GeV, the regularity of high energy photon events from different GRBs exists at a significance of 3-5~$\sigma$ with $E_{\rm LV}=3.6\times 10^{17}~\rm GeV$ determined by the GRB data.


\vspace{1cm}


\acknowledgments

This work is supported by National Natural Science Foundation of China
(Grant No.~11475006).
It is also supported by the Undergraduate Research Fund of Education Foundation of Peking University.

\clearpage

\appendix
\section{GRB Data\label{app1}}

\begin{table}[!h]
  \centering
  \renewcommand\arraystretch{1.5}
  \caption{25 bright GRBs analyzed in this artcle}
   \begin{tabular*}{0.9\textwidth}{@{\extracolsep\fill}cccccc}
     \hline
     \hline
    GRB   & $z$~\cite{redshift_data} & $t_{\rm peak}$~{\scriptsize (s)} & \multicolumn{1}{c}{GRB} & \multicolumn{1}{c}{$z$}~\cite{redshift_data} & \multicolumn{1}{c}{$t_{\rm peak}$~{\scriptsize (s)}} \\

    \hline
    160625 & 1.406 & 0.284  & \multicolumn{1}{c}{120624} & \multicolumn{1}{c}{2.1974} & \multicolumn{1}{c}{8.314 } \\

    \hline
    160509 & 1.17  & 13.920  & \multicolumn{1}{c}{110731} & \multicolumn{1}{c}{2.83} & \multicolumn{1}{c}{0.488 } \\

     \hline
    150514 & 0.807 & 0.958  & \multicolumn{1}{c}{100728} & \multicolumn{1}{c}{1.567} & \multicolumn{1}{c}{54.004 } \\
     \hline
    150403 & 2.06  & 11.388  & \multicolumn{1}{c}{100414} & \multicolumn{1}{c}{1.368} & \multicolumn{1}{c}{0.288 } \\
     \hline
    150314 & 1.758 & 1.504  & \multicolumn{1}{c}{091208} & \multicolumn{1}{c}{1.063} & \multicolumn{1}{c}{0.722 } \\
     \hline
    141028 & 2.33  & 13.248  & \multicolumn{1}{c}{091003} & \multicolumn{1}{c}{0.8969} & \multicolumn{1}{c}{5.410 } \\
     \hline
    131231 & 0.642 & 23.040  & \multicolumn{1}{c}{090926} & \multicolumn{1}{c}{2.1071} & \multicolumn{1}{c}{4.320 } \\
     \hline
    131108 & 2.40  & 0.128  & \multicolumn{1}{c}{090902} & \multicolumn{1}{c}{1.822} & \multicolumn{1}{c}{9.768 } \\
     \hline
    130702 & 0.145 & 1.788  & \multicolumn{1}{c}{090510} & \multicolumn{1}{c}{0.903} & \multicolumn{1}{c}{-0.032 } \\
    \hline
    130518 & 2.488 & 25.854  & \multicolumn{1}{c}{090328} & \multicolumn{1}{c}{0.736} & \multicolumn{1}{c}{5.378 } \\
     \hline
    130427 & 0.3399 & 0.544  & \multicolumn{1}{c}{090323} & \multicolumn{1}{c}{3.57} & \multicolumn{1}{c}{15.998 } \\
     \hline
    120729 & 0.80  & 1.488  & \multicolumn{1}{c}{080916} & \multicolumn{1}{c}{4.35} & \multicolumn{1}{c}{5.984 } \\
    \hline
    120711 & 1.405 & 69.638  &       &       &  \\
     \hline
     \hline
    \end{tabular*}%
    \begin{tablenotes}
        \item 25 bright GRBs analyzed in this article. $t_{\rm peak}$ is the first main peak time of GBM light curve, with the trigger time of GBM as the zero point.
    \end{tablenotes}
  \label{tab:grb}%
\end{table}%

\clearpage
\begin{table}[htbp]
  \centering
  \renewcommand\arraystretch{1.5}
  \caption{12 photons in data set~\uppercase\expandafter{\romannumeral4}}
    \begin{tabular*}{0.9\textwidth}{@{\extracolsep\fill}cccccc}
    \hline
     \hline
    GRB   & $z$ & $E_{\rm in}$~{\scriptsize (GeV)} & \multicolumn{1}{l}~{$\Delta t_{\rm obs}$~{\scriptsize (s)}} & $\frac{\Delta t_{\rm obs}}{1+z}$~{\scriptsize (s)} & $\kappa$~{\scriptsize ($\times 10^{18}~\mathrm{s}~\cdot$~GeV) } \\
     \hline
    160509 & 1.17  & 112.6  & 62.586 & 28.842 & 14.23  \\
     \hline
    130427 & 0.3399 & 103.3  & 18.100 & 13.509 & 9.58  \\
     \hline
    130427\footnote{\quad This photon corresponds to the hollow circle in Fig.~\ref{fig:2f}, i.e., GRB~130427$^{\star}$.} & 0.3399 & 51.8  & 77.853 & 58.104 & 4.80  \\
     \hline
    100414 & 1.368 & 70.6  & 33.081 & 13.970 & 8.73  \\
     \hline
    090926 & 2.1071 & 60.5  & 20.518 & 6.604 & 6.52  \\
     \hline
    090902 & 1.822 & 112.5  & 71.978 & 25.507 & 12.87  \\
     \hline
    090902 & 1.822 & 51.1  & 16.400 & 5.812 & 5.85  \\
     \hline
    090902 & 1.822 & 43.5  & 35.840 & 12.700 & 4.97  \\
     \hline
    090902 & 1.822 & 40.1  & 4.399 & 1.559 & 4.59  \\
     \hline
    090510 & 0.903 & 56.9  & 0.860 & 0.452 & 7.21  \\
     \hline
    080916 & 4.35  & 146.7  & 34.525 & 6.453 & 9.87  \\
     \hline
    080916 & 4.35  & 66.5  & 10.561 & 1.974 & 4.47  \\
     \hline
     \hline
    \end{tabular*}%
    \label{tab:photons_in_data_set_4}%
    \begin{tablenotes}
        \item 12 photons in data set~~\uppercase\expandafter{\romannumeral4}. $E_{\rm in}$ is the intrinsic energy of the photon at the source of GRB. $\Delta t_{\rm obs}$ is the observed time of the photon with the first main peak time of GBM light curve as zero point. $\kappa$ is defined in Eq.~(\ref{eq:kappa}).
    \end{tablenotes}
\end{table}%

\section{The choice of $\rho$ \label{app2}}
\begin{figure}[!h]
  \centering
  \includegraphics[width=0.48\textwidth]{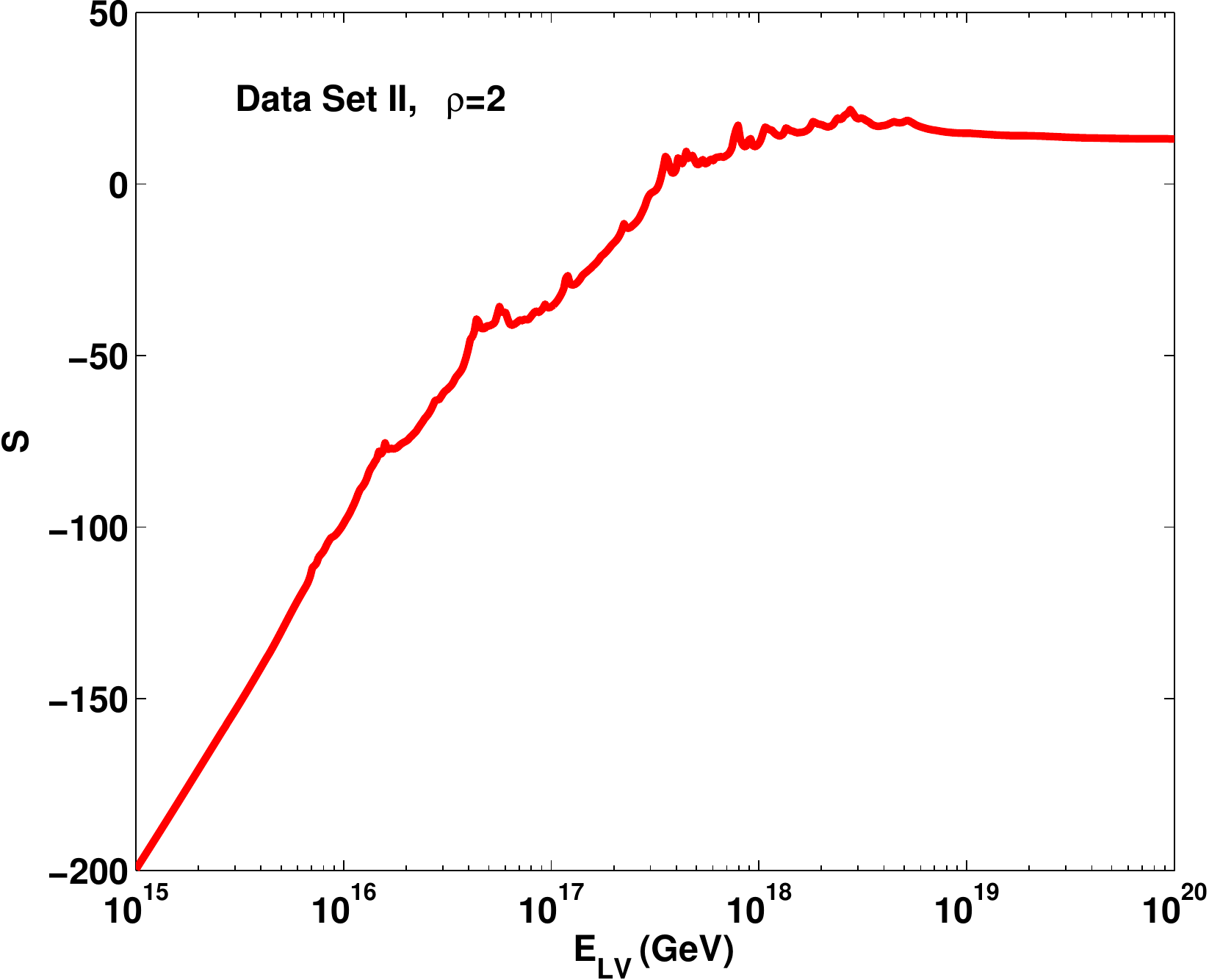}
  \caption{$\mathcal{S}$-$E_{\rm LV}$ curve for data set~\uppercase\expandafter{\romannumeral2} with $\rho=2$ in Eq.~(\ref{eq:S}). The curve fluctuates dramatically.}\label{fig:appB_rho_2}
\end{figure}

\begin{figure}
  \centering
  \includegraphics[width=0.32\textwidth]{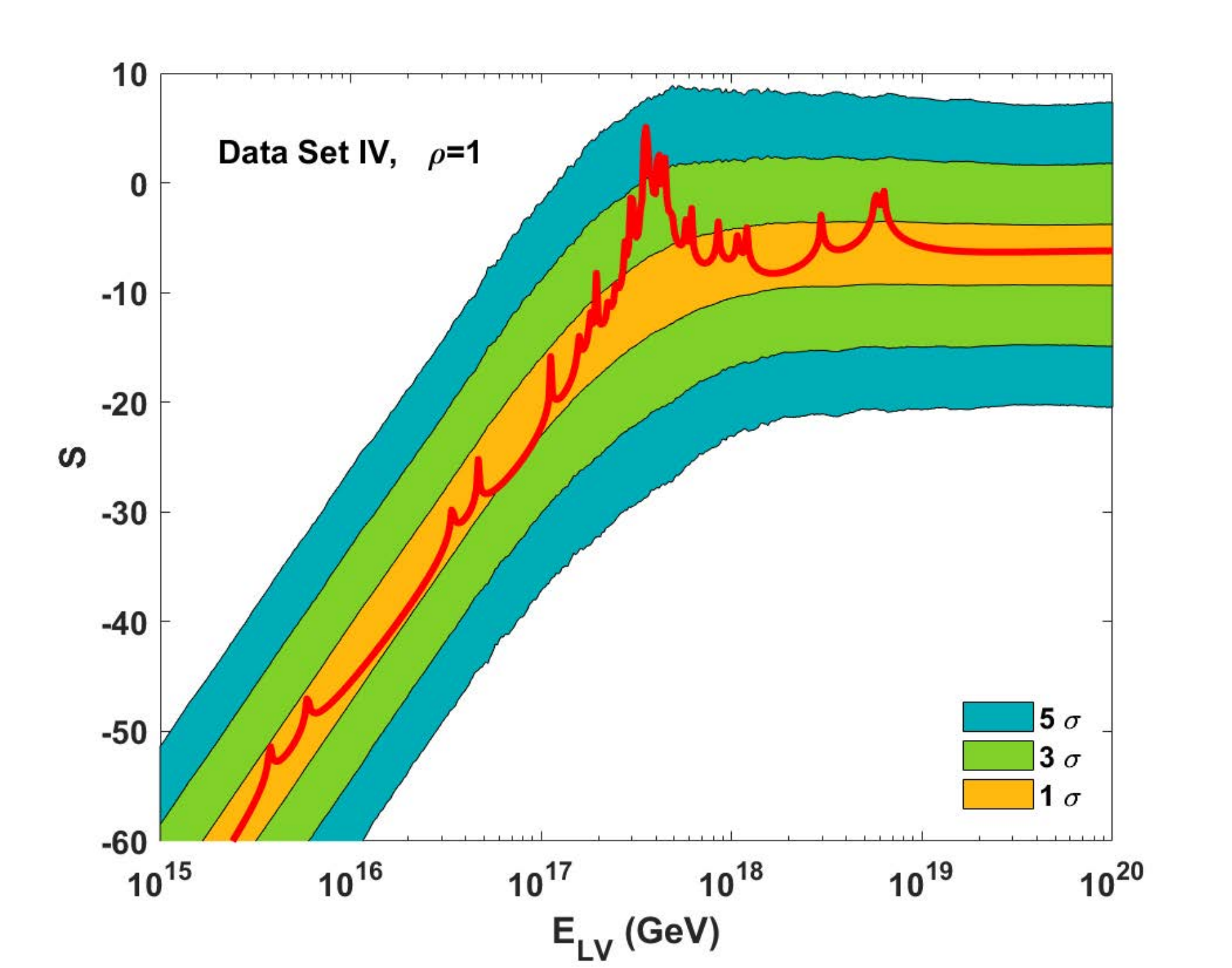}
  \includegraphics[width=0.32\textwidth]{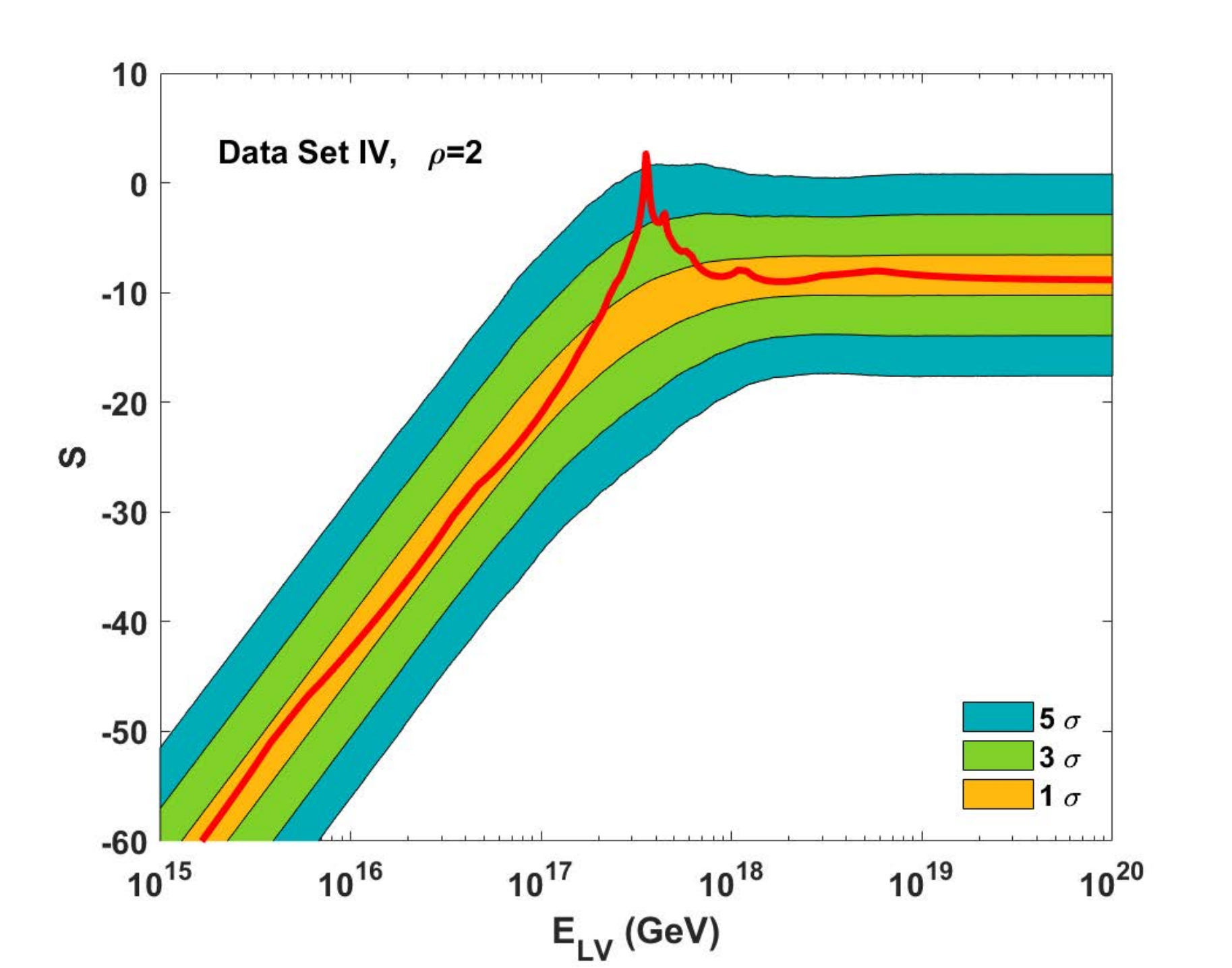}
  \includegraphics[width=0.32\textwidth]{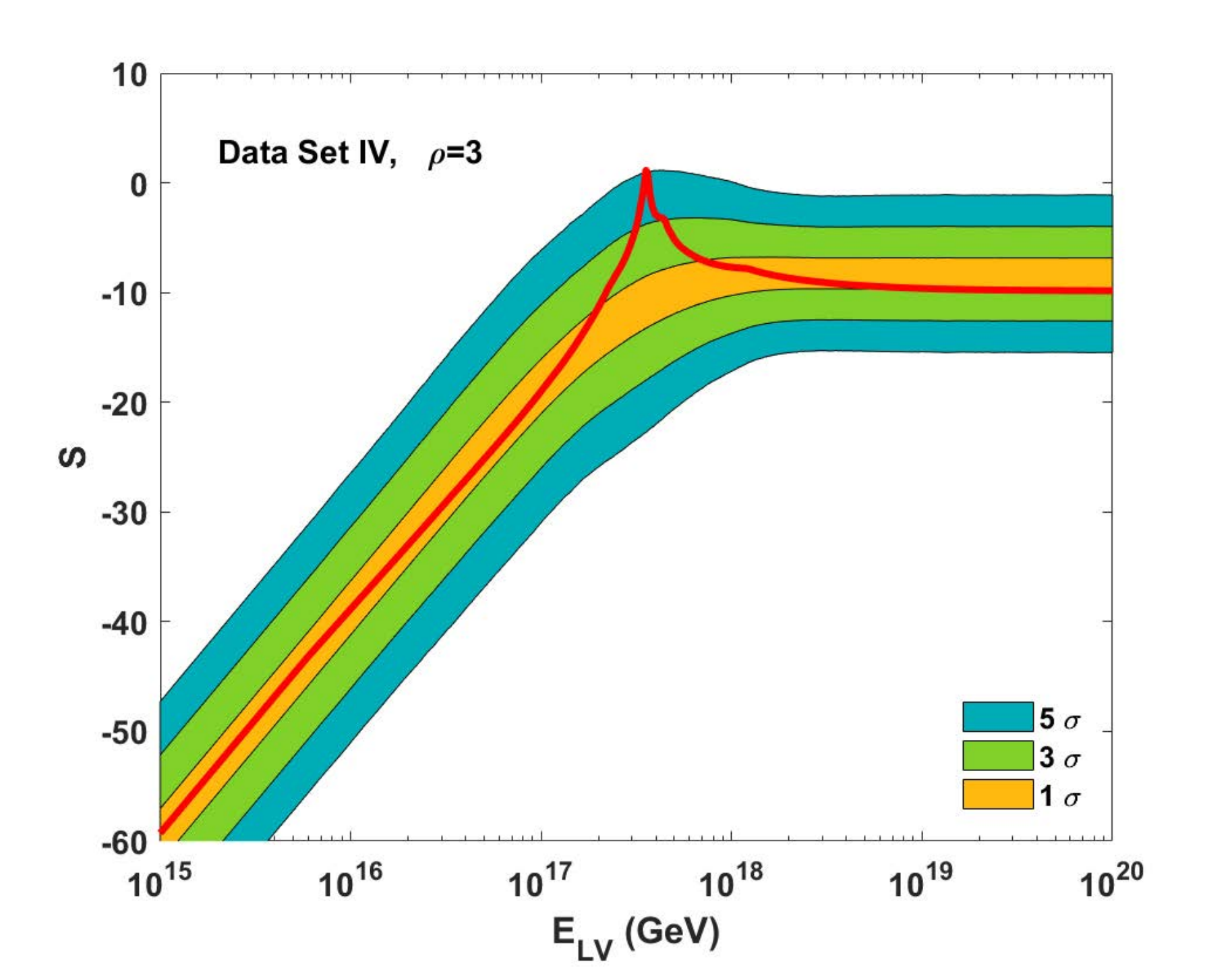}\\
  \includegraphics[width=0.28\textwidth]{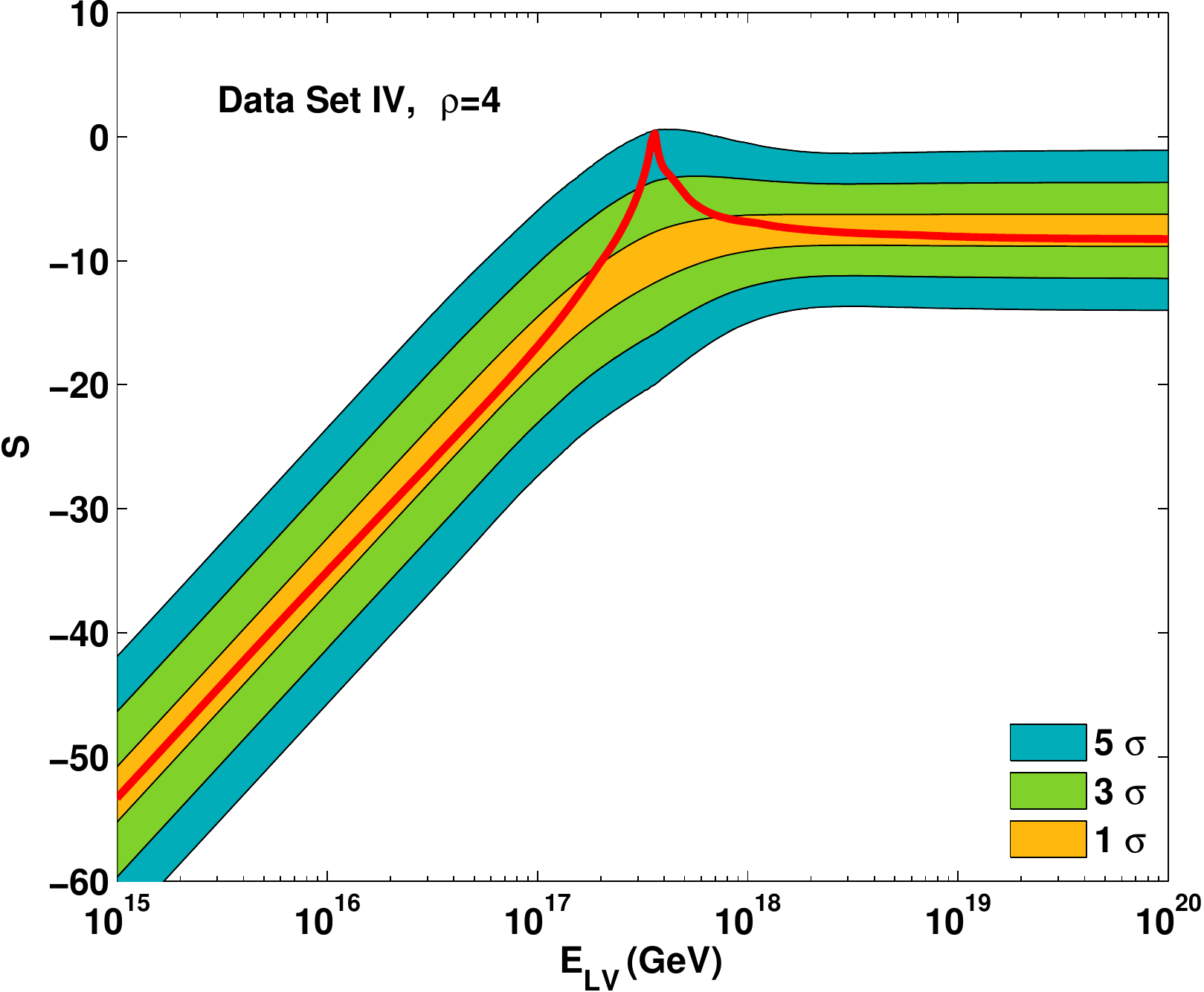}
  \includegraphics[width=0.28\textwidth]{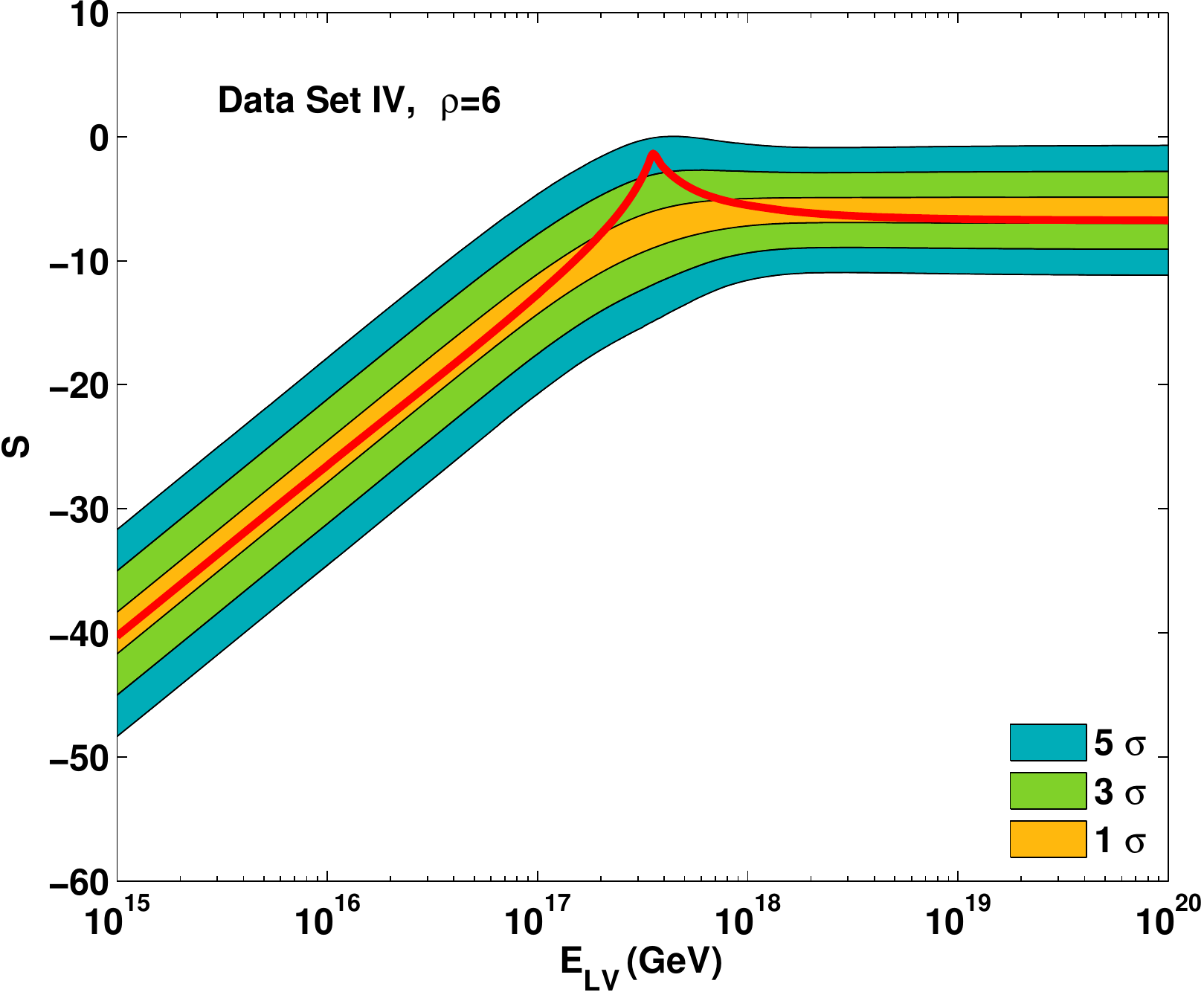}
  \caption{$\mathcal{S}$-$E_{\rm LV}$ curve for data set~\uppercase\expandafter{\romannumeral4} with $\rho=1\sim4$ and 6 in Eq.~(\ref{eq:S}) respectively. }\label{fig:appB_rho_4_6}
\end{figure}

The smallest data set in our analysis, data set~\mbox{\uppercase\expandafter{\romannumeral4}}, contains only 12 photon events. Therefore we cannot choose a $\rho$ greater than 6. The reason is obvious. According to Eq.~(\mbox{\ref{eq:S}}), the 8-th photon event will be excluded from the analysis if $\rho=7$. A too small $\rho$, say $\rho=2$, is also not reasonable since the $\mathcal{S}$-$E_{\rm LV}$ curve would fluctuate dramatically. An example for data set~\uppercase\expandafter{\romannumeral2} is shown in Fig.~\mbox{\ref{fig:appB_rho_2}}. After careful examination, we find that $\rho=4$, 5 or 6 is reasonable and $\rho$ is set as 5 in the main body of this article.
The $\mathcal{S}$-$E_{\rm LV}$ curves for data set~\mbox{\uppercase\expandafter{\romannumeral4}} with alternative $\rho$ are shown in Fig.~\mbox{\ref{fig:appB_rho_4_6}}. We can see that the curve is more flat when $\rho$ is 6. But the regions of random data are also more flat. Thus the conclusion is similar: the peak of the $\mathcal{S}$-$E_{\rm LV}$ curve for observed data goes beyond the $3\, \sigma$ region of random data but is still within the $5\, \sigma$ region.

\section{The test of the method \label{app3}}
In order to test the validity of our method, it is applied to some artificially produced photon events. In Test~\uppercase\expandafter{\romannumeral1}, an $E_{\rm LV} = 4 \times 10^{17}~\rm GeV$ is planted to photons emitted at the same time from the GRB source with redshift $z = 1$, i.e., $\Delta t_{\rm in} = 25~\rm s$ in Eq.~(\ref{eq:intrinsic_lag}). In Test~\uppercase\expandafter{\romannumeral2}, a fictitious $E_{\rm LV}$ is still set as $4 \times 10^{17}~\rm GeV$. But photons have $\Delta t_{\rm in}$ uniformly distributed between $0 \sim 50~\rm s$. In Test~\uppercase\expandafter{\romannumeral3}, $E_{\rm LV}$ is no longer set as a fixed number, but has a uniform distribution between $10^{17} \sim 10^{18}~\rm GeV$ for different photons. In all tests, the energy of photons has a uniform distribution between $0\sim 100~\rm GeV$.

\begin{figure}
  \centering
  \includegraphics[width=0.48\textwidth]{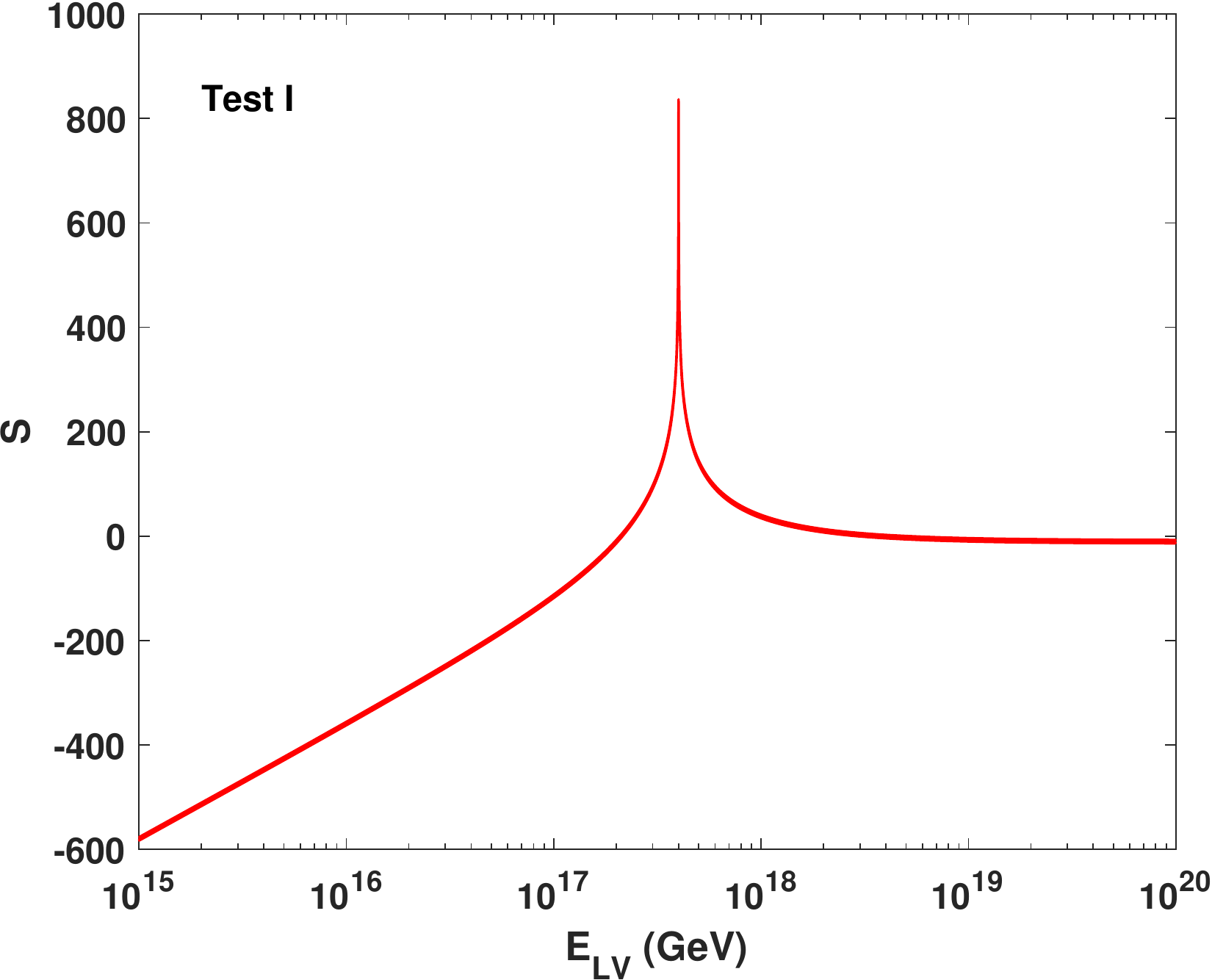}
  \includegraphics[width=0.48\textwidth]{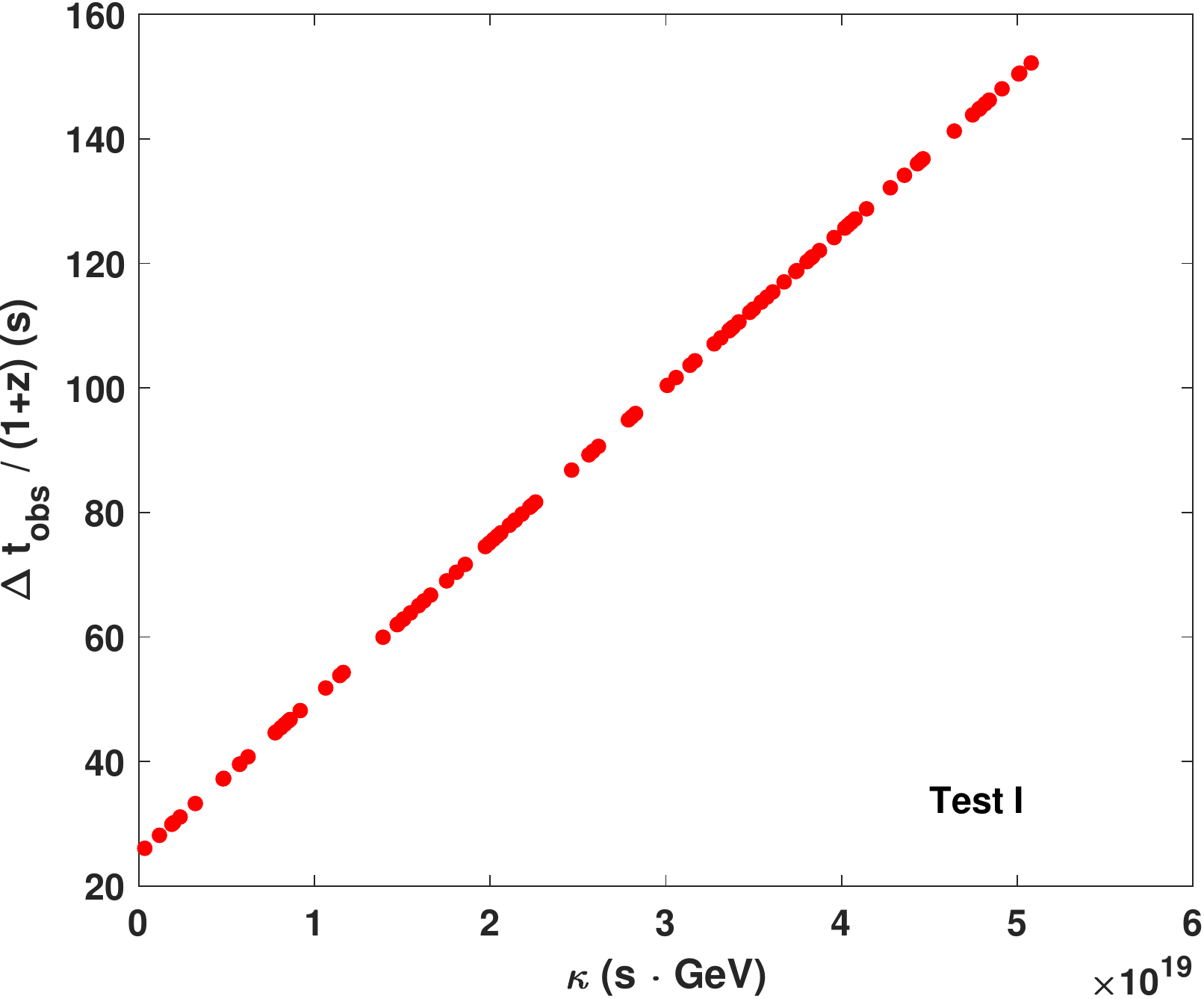}\\
  \includegraphics[width=0.48\textwidth]{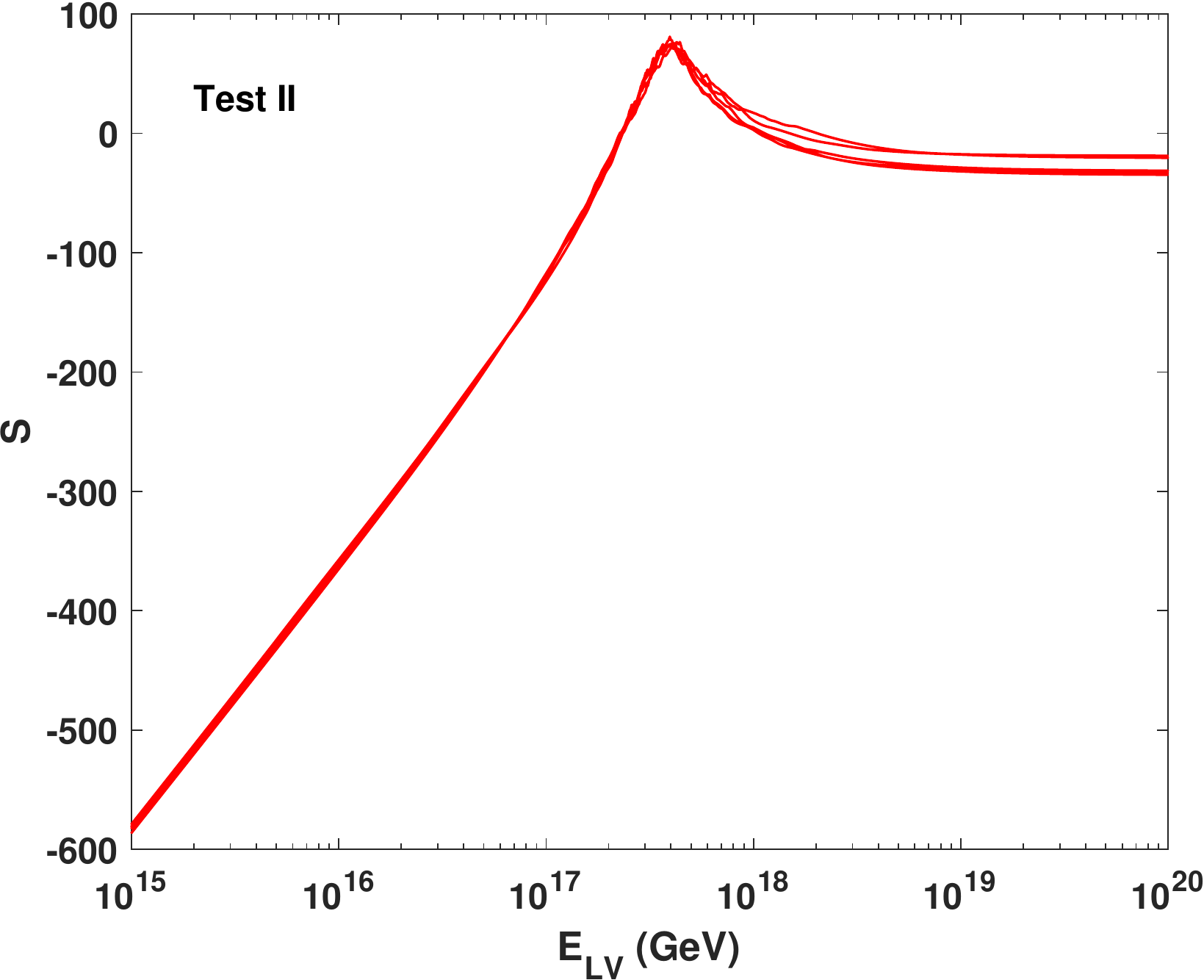}
  \includegraphics[width=0.48\textwidth]{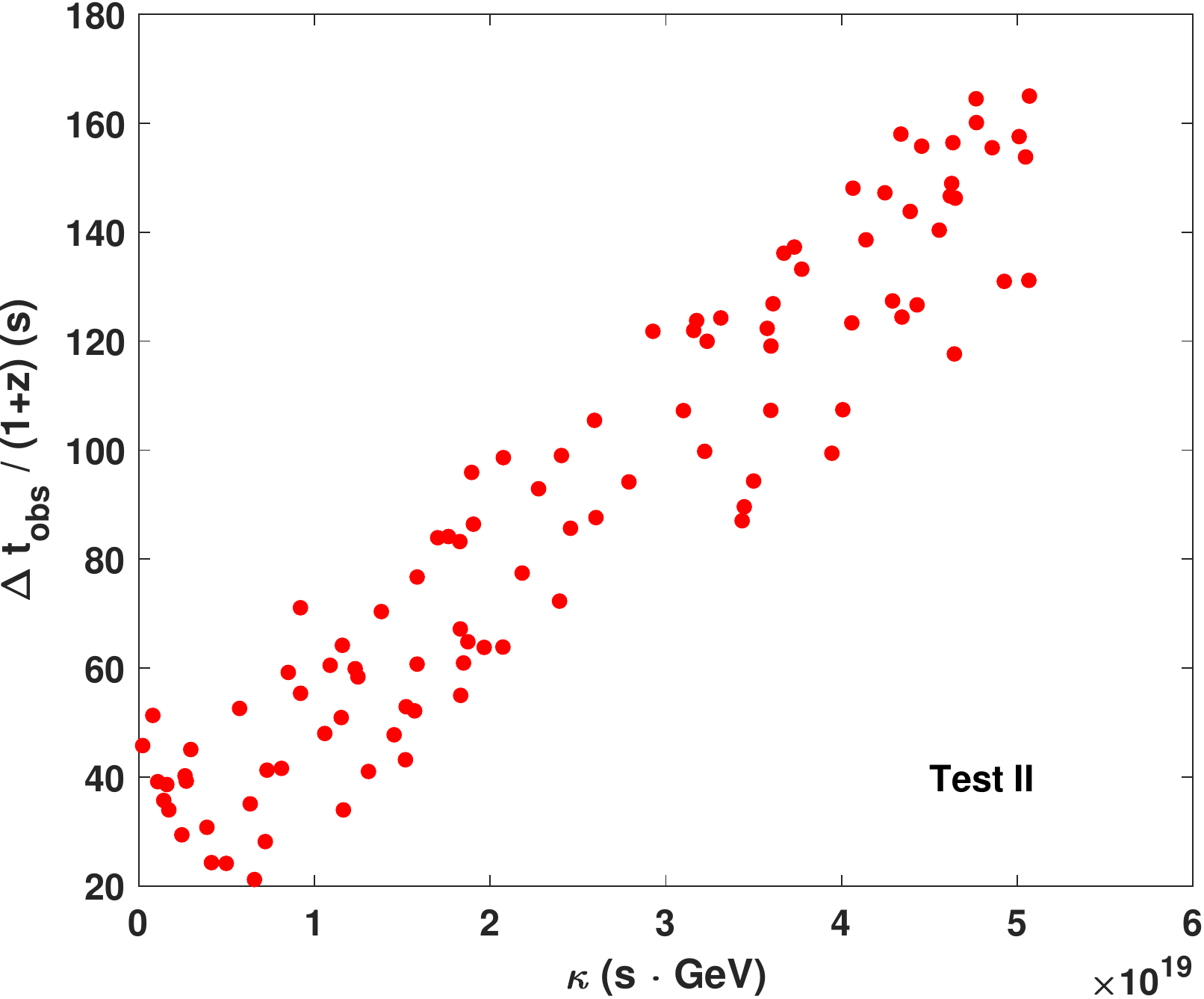}\\
  \includegraphics[width=0.48\textwidth]{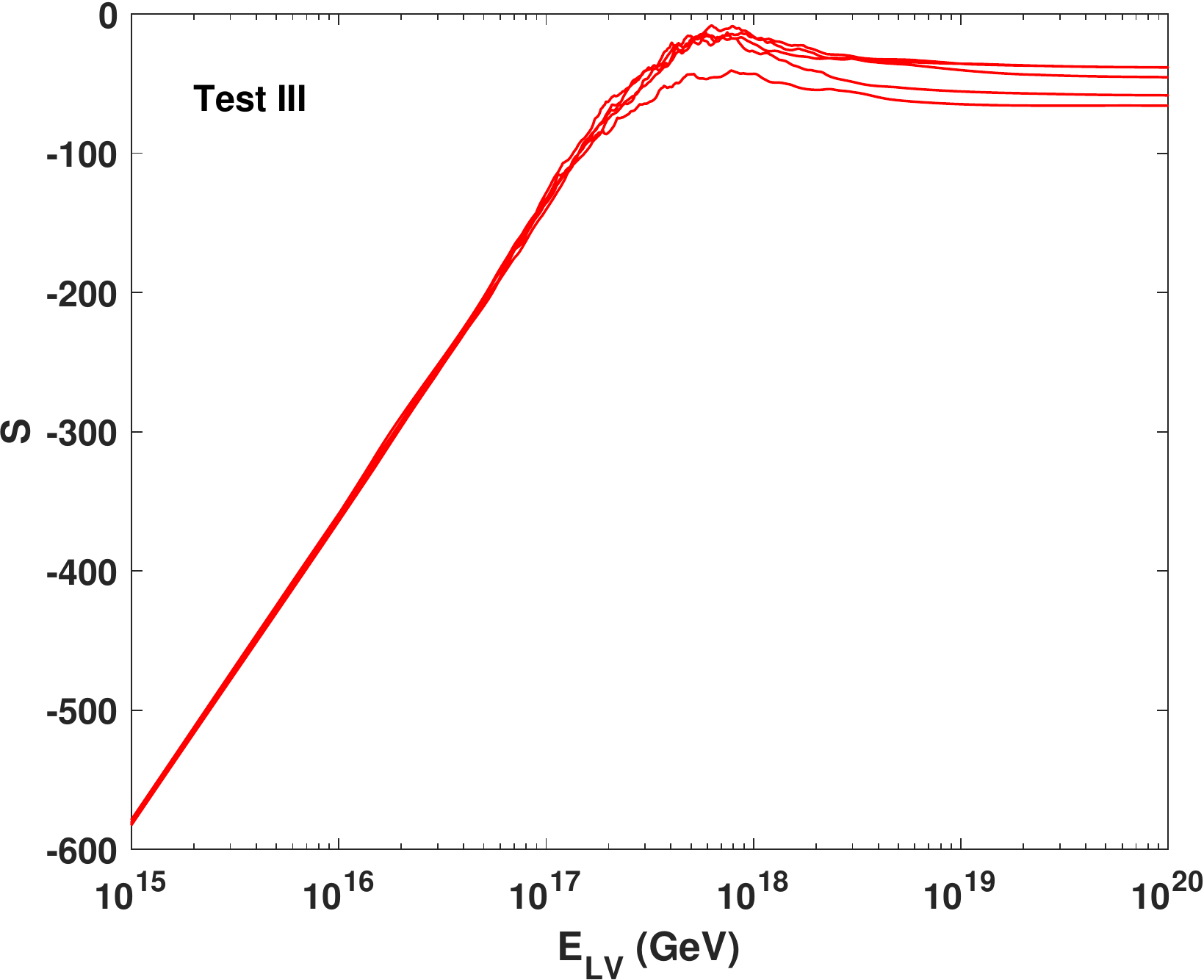}
  \includegraphics[width=0.48\textwidth]{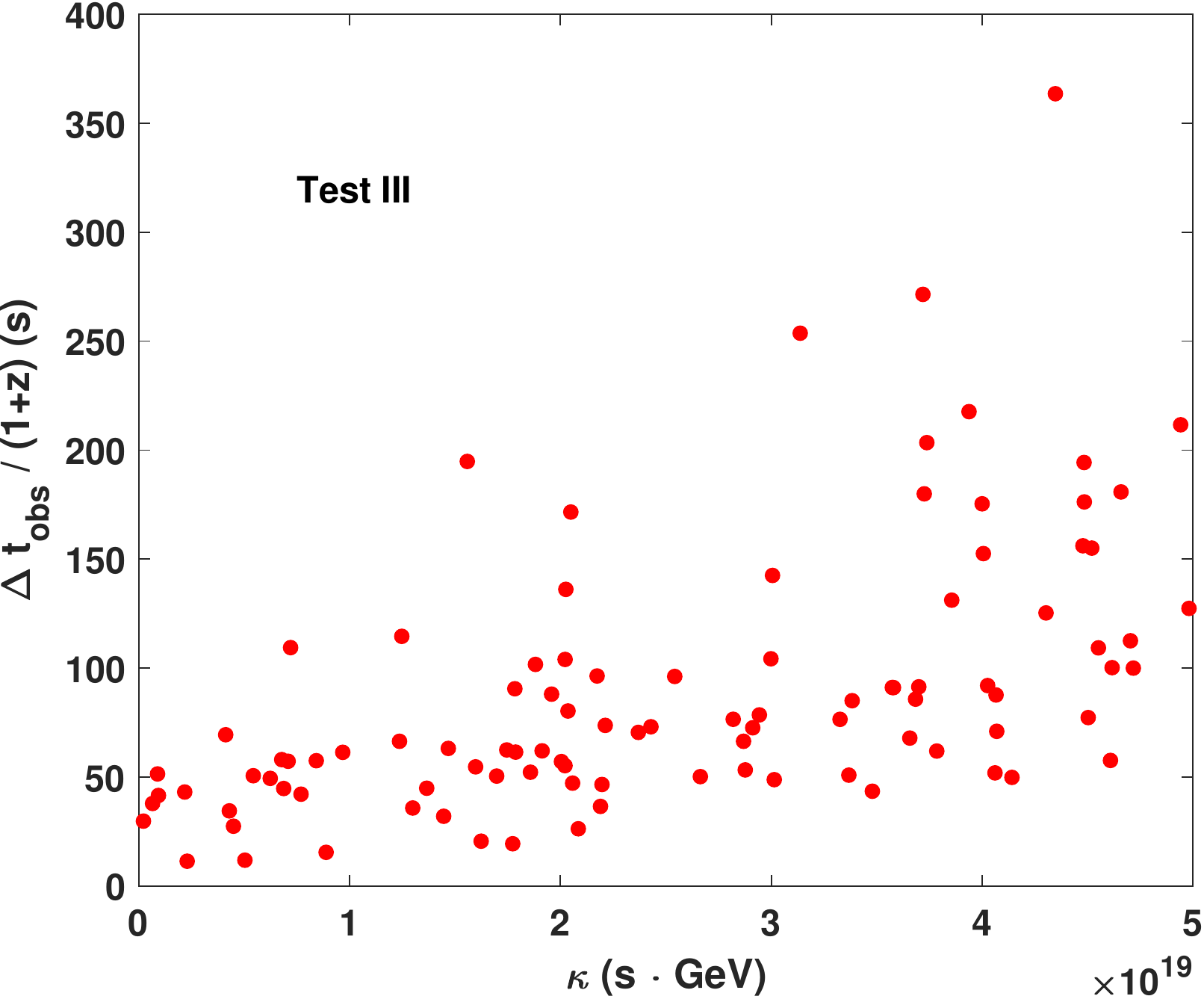}\\
  \caption{Typical results for Tests~\uppercase\expandafter{\romannumeral1} $\sim$~\uppercase\expandafter{\romannumeral3} and data sets with $N = 100$. The results are in principle approximately identical for different simulations for the left panels, whereas the right panels correspond to the case of one simulation for each test. }\label{fig:test_N_100}
\end{figure}

\begin{figure}
  \centering
  \includegraphics[width=0.48\textwidth]{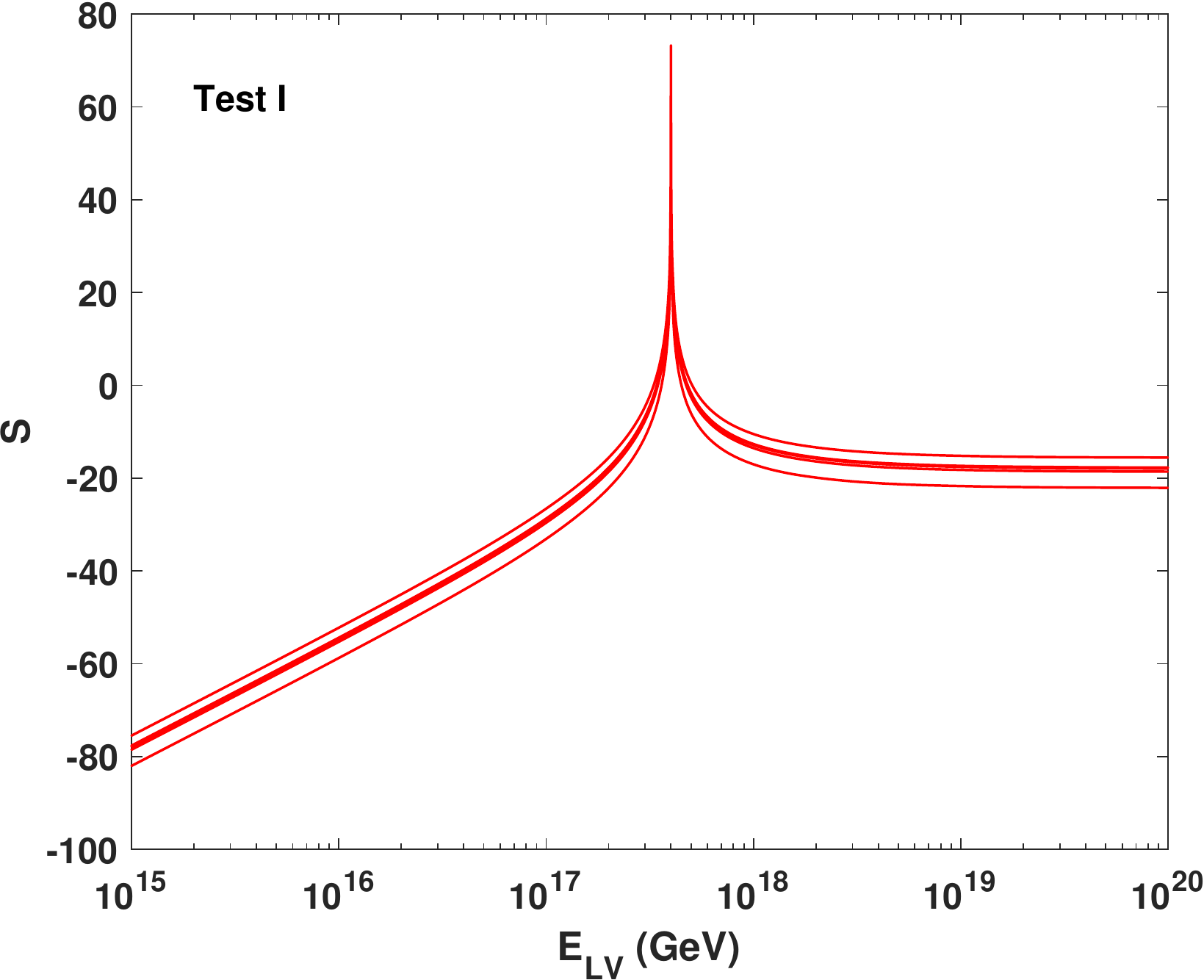}
  \includegraphics[width=0.48\textwidth]{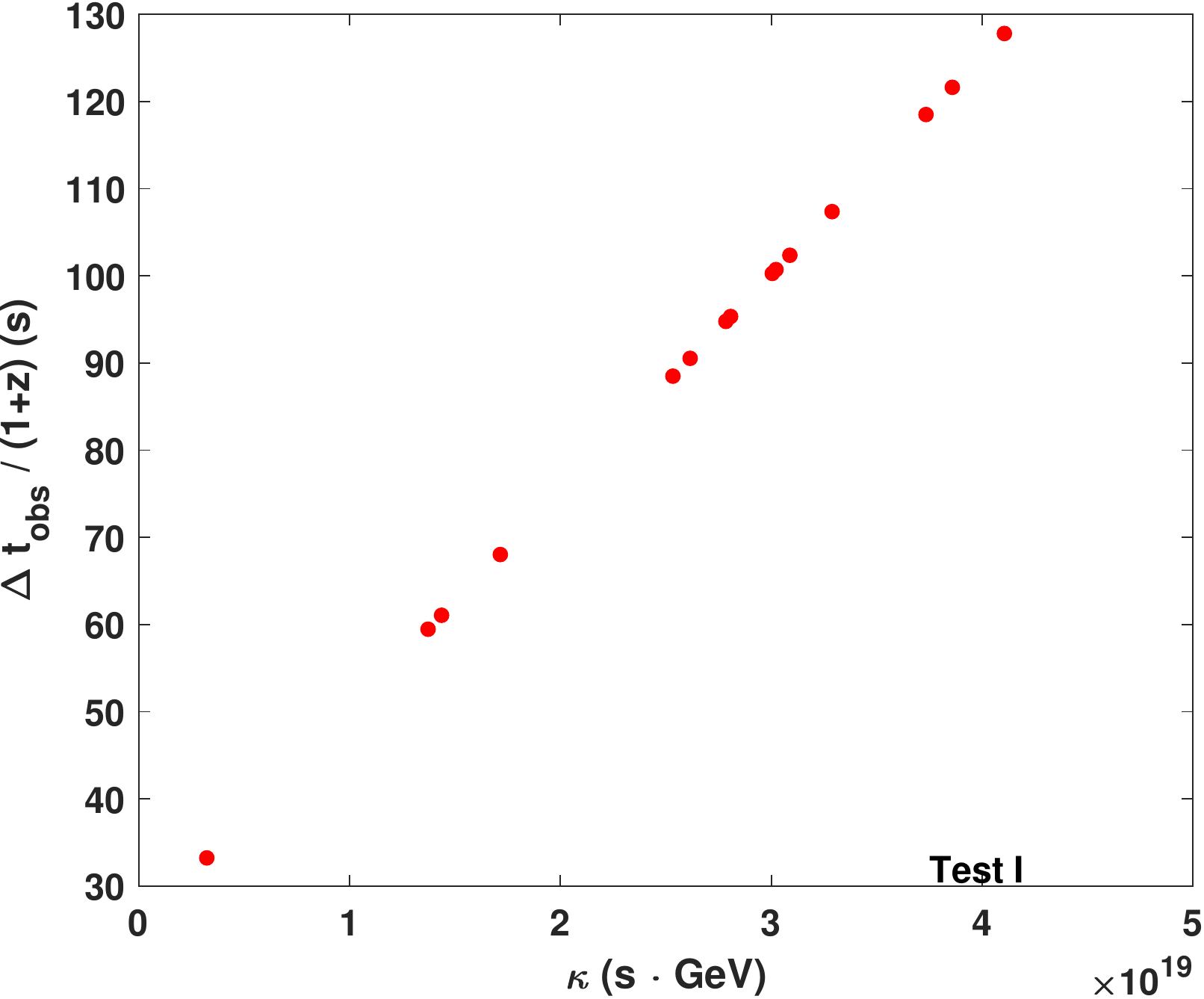}\\
  \includegraphics[width=0.48\textwidth]{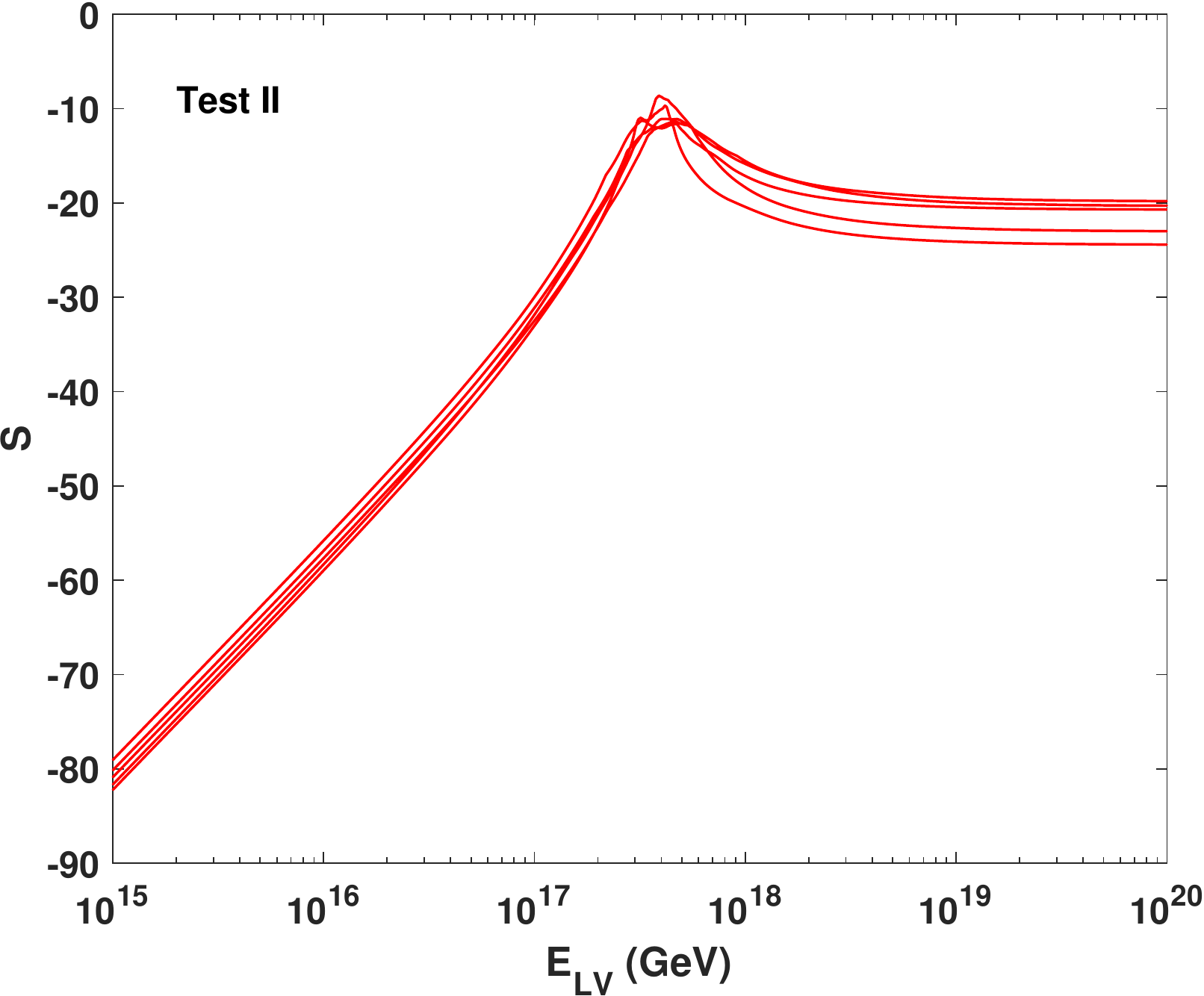}
  \includegraphics[width=0.48\textwidth]{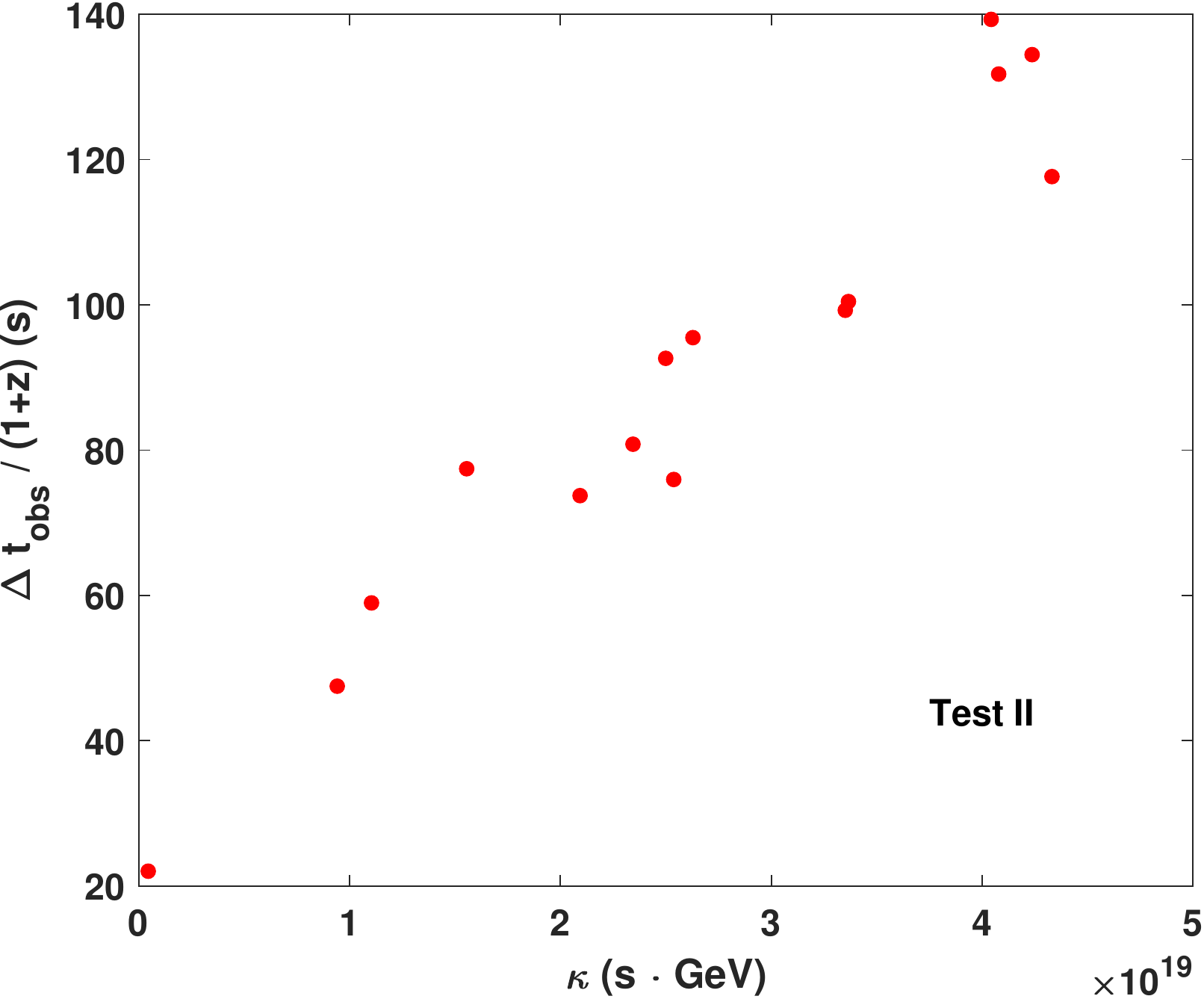}\\
  \includegraphics[width=0.48\textwidth]{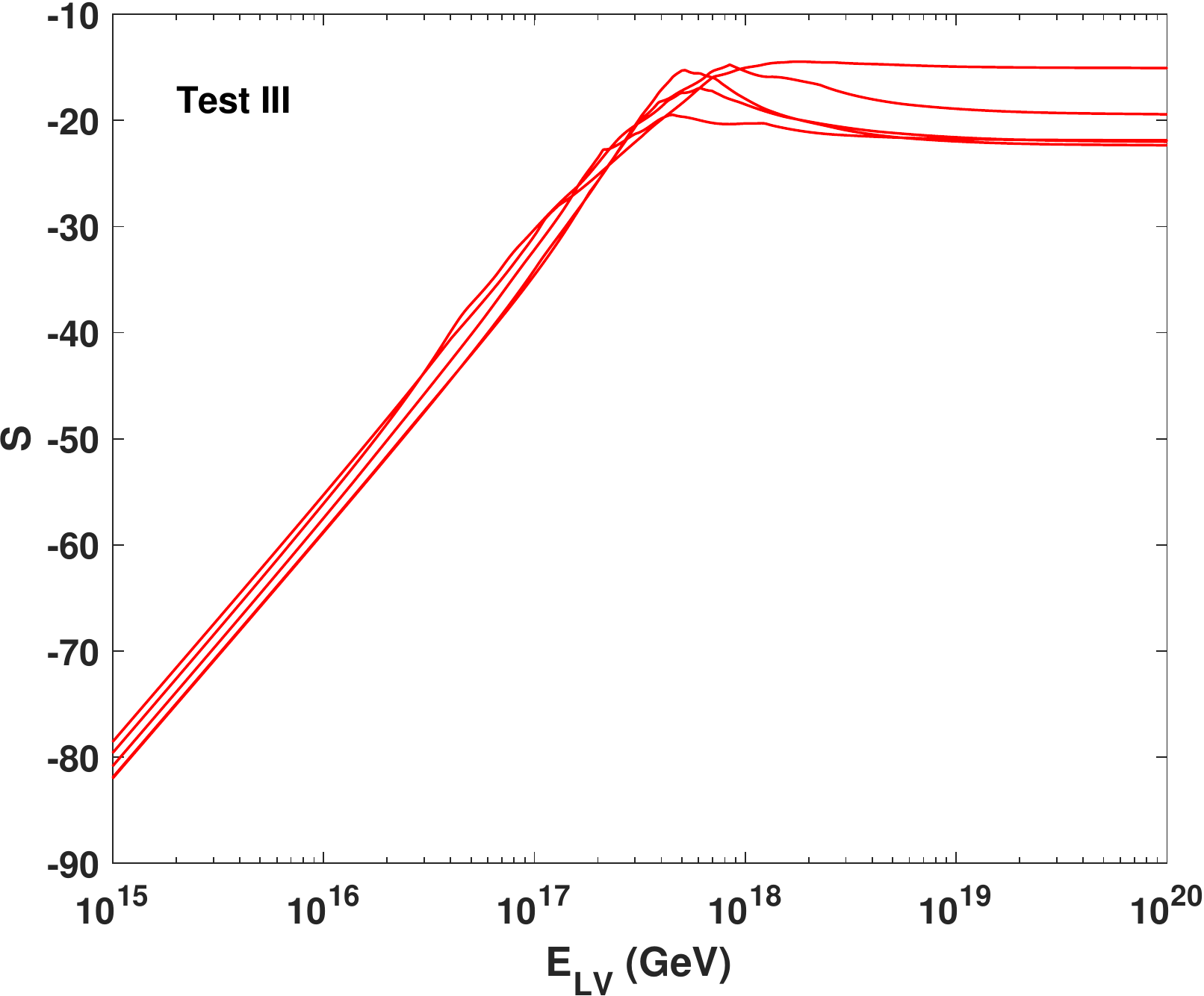}
  \includegraphics[width=0.48\textwidth]{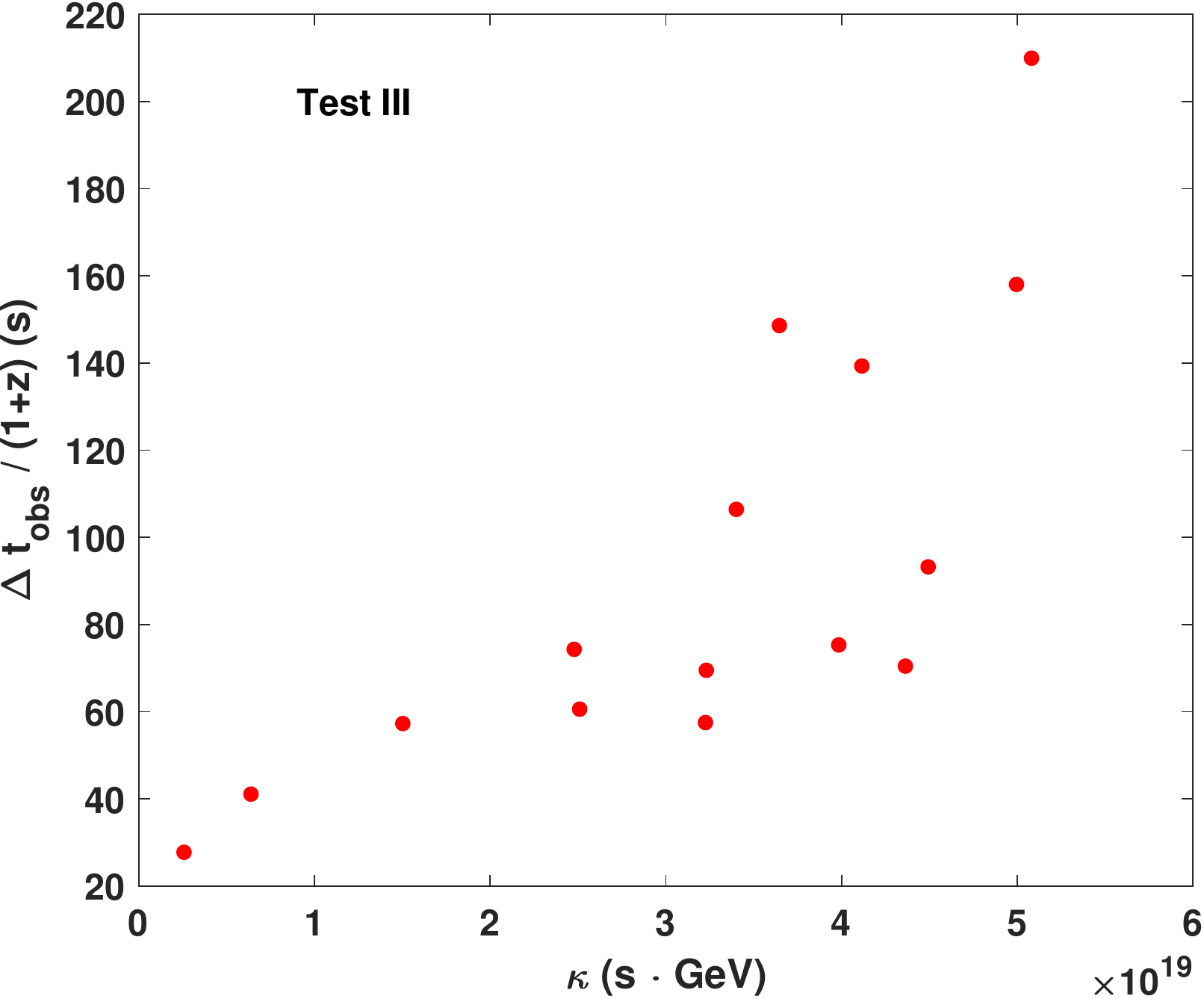}\\
  \caption{Typical results for Tests~\uppercase\expandafter{\romannumeral1} $\sim$~\uppercase\expandafter{\romannumeral3} and data sets with $N = 15$.}\label{fig:test_N_15}
\end{figure}

Tests~\uppercase\expandafter{\romannumeral1} $\sim$~\uppercase\expandafter{\romannumeral3} are applied to two data sets, with $N = 100$ and 15 randomly produced photons respectively. Such simulations are repeated for $10^{3}$ times. Results of five typical simulation are shown in Figs.~\ref{fig:test_N_100}, \ref{fig:test_N_15}. $\mathcal{S}$-$E_{\rm LV}$  curves are shown in the left panels. We can see that for Test~\uppercase\expandafter{\romannumeral1}, there is a very significant peak at $E_{\rm LV} = 4 \times 10^{17}~\rm GeV$, identical for all simulations. For Test~\uppercase\expandafter{\romannumeral2}, there is also a peak at approximately $E_{\rm LV} = 4 \times 10^{17}~\rm GeV$, but less significant than the Test~\uppercase\expandafter{\romannumeral1} case. In some simulations, the position of the peak is not accurate for the $N = 15$ data set. For Test~\uppercase\expandafter{\romannumeral3}, since $E_{\rm LV}$ has a uniform distribution, only very faint peaks can be seen in only some of the simulations and the positions of these peaks are different and ranges between $10^{17} \sim 10^{18}~\rm GeV$, as expected.

In the right panels, the~$\Delta t_{\rm obs}/(1+z)$-$\kappa$~plots for one simulation are drawn as a reference.

\clearpage
\begin{thebibliography}{99}

\bibitem{lrr-2005-5}
For a review, see, e.g., D.~Mattingly,
Modern tests of Lorentz invariance,
{Living Rev. Relativ.} {8} (2005) 5.


\bibitem{method1}
G.~Amelino-Camelia {\it et al.},
Distance measurement and wave dispersion in a Liouville-string approach to quantum gravity,
{Int.\ J.\ Mod.\ Phys.\ A} {12} (1997) 607.

\bibitem{method2}
G.~Amelino-Camelia, J.~Ellis, N.~E.~Mavromatos, {\it et al.},
Tests of quantum gravity from observations of gamma-ray bursts,
{Nature} {393} (1998) 763.

\bibitem{oldformula}
J.~Ellis, N.~E.~Mavromatos, D.~V.~Nanopoulos, A.~S.~Sakharov,
Quantum-gravity analysis of gamma-ray bursts using wavelets,
{Astron.\ Astrophys.\ } {402} (2003) 409-424.

\bibitem{Ellis_app}
J.~R.~Ellis, N.~E.~Mavromatos, D.~Nanopoulos, A.~S.~Sakharov, E.~K.~G.~Sarkisyan,
Robust limits on Lorentz violation from gamma-ray bursts,
{Astropart.\ Phys.\ } {25} (2006) 402-411. [Corrigendum {29} (2008) 158-159].

\bibitem{Rodri}
M.~Rodriguez~Martinez, T.~Piran, and Y.~Oren,
GRB 051221A and tests of Lorentz symmetry,
{J. Cosmol. Astropart. Phys.} {5} (2006) 017.


\bibitem{Lamon_grg}
R.~Lamon, N.~Produit, and F.~Steiner,
Study of Lorentz violation in INTEGRAL gamma-ray bursts,
{Gen. Relativ. Gravit.} {40} (2008) 1731.

\bibitem{DisCan}
J.~D.~Scargle, J.~P.~Norris, and J.~T.~Bonnell,
An algorithm for detecting quantum gravity photon dispersion in gamma-ray bursts: DisCan,
{Astrophys. J.} {673} (2008) 972.

\bibitem{Abdo_1}
A.~A.~Abdo, M.~Ackermann, M.~Arimoto, K.~Asano, W.~B.~Atwood {\it et al.},
Fermi observations of high-energy gamma-ray emission from GRB 080916C,
{Science} {323} (2009) 1688.

\bibitem{Abdo_2}
A.~A.~Abdo, M.~Ackermann, M.~Ajello, K.~Asano, W.~B.~Atwood {\it et al.},
A limit on the variation of the speed of light arising from quantum gravity effects,
{Nature} {462} (2009) 331.

\bibitem{Xiao:2009xe}
  Z.~Xiao and B.-Q.~Ma,
Constraints on Lorentz invariance violation from gamma-ray burst GRB090510,
  {Phys.\ Rev.\ D} {80} (2009) 116005.



\bibitem{shaolijing}
L.~Shao, Z.~Xiao, B.-Q.~Ma,
Lorentz violation from cosmological objects with very high energy photon emissions,
{Astropart.\ Phys.\ } {33} (2010) 312-315.

\bibitem{SMM}
V.~Vasileiou, A.~Jacholkowska, F.~Piron, {\it et al.},
Constraints on Lorentz invariance violation from Fermi-Large Area Telescope observations of gamma-ray bursts,
{Phys. Rev. D} {87} (2013) 122001.


\bibitem{zhangshu}
S.~Zhang, B.-Q.~Ma,
Lorentz violation from gamma-ray bursts,
{Astropart.\ Phys.\ } {61} (2015) 108-112.

\bibitem{Vlasios Vasileiou}
V.~Vasileiou, J.~Granot, T.~Piran, G.~Amelino-Camelia,\
A Planck-scale limit on spacetime fuzziness and stochastic Lorentz invariance violation,
{Nature \ Physics} {11} (2015) 344.

\bibitem{Xu_app}
H.~Xu, B.-Q.~Ma,
Light speed variation from gamma-ray bursts,
{Astropart.\ Phys.\ } {82} (2016) 72.

\bibitem{Xu_plb}
H.~Xu and B.-Q.~Ma,
Light speed variation from gamma ray burst GRB160509A,
{Phys. Lett. B} {760} (2016) 602.

\bibitem{note-added}
G.~Amelino-Camelia, G.~D'Amico, G.~Rosati and N.~Loret,
In-vacuo-dispersion features for GRB neutrinos and photons,
Nature Astronomy 1 (2017) 0139.


\bibitem{LAT}
W.~B.~Atwood, A.~A.~Abdo, M.~Ackermann, {\it et al.},
The Large Area Telescope on the Fermi Gamma-ray Space Telescope mission,
{Astrophys.\ J.\ } {697} (2009) 1071-1102.


\bibitem{GBM}
C.~Meegan, G.~Lichti, P.~N.~Bhat {\it et al.},
The Fermi Gamma-Ray Busrt Monitor,
{Astrophys.\ J.\ } {702} (2009) 791.


\bibitem{newformula}
U.~Jacob, T.~Piran,
Lorentz-violation-induced arrival delays of cosmological particles,
{JCAP}  {0801} (2008) 031.

\bibitem{pgb}
K.~A.~Olive, {\it et al.}, (Particle Data Group),
Review of particle physics,
{Chin. Phys. C }  {38} (2014) 090001.


\bibitem{information_entropy}
C.~E.~Shannon,
A mathematical theory of communication,
{Bell Syst. Tech. J.}, {27(3)} (1948) 379-423.


\bibitem{minimum_disp}
J.~Ellis, N.~Harries, A.~Meregaglia, A.~Rubbia, and A.~S.~Sakharov,
Probes of Lorentz violation in neutrino propagation,
{Phys. Rev. D} {78} (2008) 033013.

\bibitem{energy_cost_fun}
J.~Albert, {\it et al.} [MAGIC Collaboration] and J.~Ellis, N.~E.~Mavromatos, D.~V.~Nanopoulos, {\it et al.},
Probing quantum gravity using photons from a flare of the active galactic nucleus Markarian 501 observed by the MAGIC telescope,
   {Phys.\ Lett.\ B} {668} (2008) 253.




\bibitem{LATdata}
Specific data can be downloaded from \url{http://fermi.gsfc.nasa.gov/ssc/data/access/} or \url{http://fermi.gsfc.nasa.gov/ssc/observations/types/grbs/lat_grbs/table.php}.



\bibitem{redshift_data}
Information about redshift can be found in GCN Circular Archive \url{http://gcn.gsfc.nasa.gov/gcn3_archive.html} or \url{http://www.mpe.mpg.de/~jcg/grbgen.html}.


\end{thebibliography}
\end{document}